\documentclass[11pt,a4paper]{JHEP3}
\usepackage{amsmath, amssymb}
\usepackage{verbatim}
\usepackage{latexsym}

\def\be{\begin{eqnarray}}
 \def\ee{\end{eqnarray}}
 \def\0{\nonumber}
\def\d{\partial}
\usepackage{euscript}

\usepackage{amsfonts}
\usepackage{verbatim}

\newcommand\E{{\cal E}}

\def\tphi{\tilde\phi}
\def\del{\partial}

\def\del{\partial}

\def\no{\noindent}

\def\cos{{\rm cos}}
\def\sin{{\rm sin}}

\def\bra#1{\langle #1 |}
\def\ket#1{|#1 \rangle}
\def\0{\nonumber}

\preprint{SISSA/61/2010/EP\\ ULB-TH/10-30\\\tt hep-th/1009.4158}

\title{Relevant Deformations in Open String Field Theory:\\ a Simple Solution for Lumps}

\author{ L.Bonora\\
International School for Advanced Studies (SISSA)\\
Via Bonomea 165, 34136 Trieste, Italy, and INFN, Sezione di
Trieste\\
E-mail:   \email{bonora@sissa.it},}

\author{C.Maccaferri\\
International Solvay Institutes and
Physique Th\'eorique et Math\'ematique,\\
ULB C.P. 231, Universit\'e Libre
de Bruxelles,  B-1050, Bruxelles, Belgium\\
E-mail:  \email{maccafer@gmail.com},}

\author{D.D.Tolla\\
Department of Physics and University College,
Sungkyunkwan University,
Suwon 440-746, South Korea\\
E-mail:  \email{ddtolla@skku.edu}}

\abstract{We propose a remarkably simple solution of cubic open string field theory which describes inhomogeneous tachyon condensation. The solution is in one--to-one correspondence with
the IR fixed point of the RG--flow generated in the two--dimensional world--sheet theory by integrating a relevant operator with mild enough OPE on the boundary. It is shown how the closed string overlap
correctly captures the shift in the closed string one point function between the UV and the IR limits of the flow. Examples of lumps in non-compact and compact transverse directions are given. }

\keywords{String Field Theory}

\begin{document}
 \maketitle
\section{Introduction and outlook}

It is expected that  Witten's covariant cubic string field theory,
\cite{Witten:1985cc}, has classical solutions  describing Boundary
Conformal Field Theories (BCFT) which are different from the one we
use to  define it. This expectation has been checked by the explicit
construction of analytic solutions describing the tachyon
vacuum~\cite{Schnabl05, Okawa1, ErlerSchnabl, RZ06, ORZ, Fuchs0,
Erler:2006hw, Erler:2006ww,  Erler:2007xt, Arroyo:2010fq,
Zeze:2010jv, Zeze:2010sr, Arroyo:2010sy} and general marginal
boundary deformations of the initial BCFT~\cite{KORZ,
Schnabl:2007az, Kiermaier:2007vu, Fuchs3, Lee:2007ns, Kwon:2008ap,
Okawa2, Okawa3, Kiermaier:2007ki, Erler:2007rh}. The well known
worldsheet description of such systems is now {\it analytically}
understood  in the context of cubic OSFT (see \cite{Fuchs4,
Schnabl:2010tb} for review). This is mainly due to the progress made
in identifying very simple (differential) subalgebras of the
$*$--product, such as the one of wedge states with
insertions~\cite{Rastelli:2000iu, Schnabl:2002gg,
Bonora:2007tm, Bonora:2009he, Bonora:2009hf}.

There are other situations which are pretty well understood  from a purely
 worldsheet perspective and we would like to be able to describe within
 OSFT: relevant boundary deformations of the initial BCFT. While marginal deformations
 describe the moduli space of the theory around the perturbative vacuum
 (thus deforming the worldsheet field theory while preserving conformal invariance),
  relevant deformations are  drastically different  in that they correspond to unstable
   directions in the string--field potential which might (or might not) lead to a new
    vacuum with less space--time energy than the original one. On the worldsheet,
    they break conformal invariance at the boundary and thus trigger an RG flow to a
    new BCFT, which has the same bulk content, but different boundary conditions. Moreover,
    the loss of space--time energy is encoded in the boundary entropy (the log of the disk partition
    function), which according to the $g$--theorem/conjecture decreases along the RG flow~\cite{Affleck}.
    Actually the tachyon vacuum is the simplest of such RG flows as it corresponds to a trivial BCFT where
    all the boundary degrees of freedom have been killed. This IR fixed point is achieved by integrating
    over the worldsheet boundary a constant (positive) tachyon vertex operator. Different, less trivial,
    IR fixed points can be reached by giving a profile to the tachyon. In this case one expects to get
    lower dimensional D--branes, placed at the minima of the boundary world-sheet potential (which is
    just the tachyon profile) \cite{Harvey, Elitzur}.

With this picture in mind, we will propose a very simple solution
of the OSFT equation of motion which is meant to describe the IR
fixed point of a relevant boundary deformation of the initial
$BCFT_0$, where the tachyon (which is the only matter field which
can be relevant) is allowed to have a non trivial space--time
dependence.
There has been evidence since the seminal work of Moeller, Sen and Zwiebach, \cite{lumps}, where a lower dimensional
D--brane has been seen to emerge as  inhomogeneous tachyon condensation in the
 modified level truncation scheme in Siegel gauge, that these solutions should exist. However an analytic expression for them (in a suitable gauge) has not yet been found.\\
Here, inspired by the original proposal of Ellwood \cite{Ellwood:2009zf},  we give a concrete construction for such a solution.
It should be stressed, however, that (differently from the numerical analysis of \cite{lumps}) we do not provide a way to {\it solve} the RG flow using OSFT but, less ambitiously, we
explicitly show that whenever the RG flow is known on the worldsheet, a
simple solution of OSFT can be written down which represents the IR fixed point
 of the flow, reproducing the correct coupling with on-shell closed strings. While, at this stage, this does not immediately give new tools to explore unknown regions of the open string landscape, we believe it is
still important in keeping with the paradigm that OSFT is a consistent and complete space-time theory
 that  {\it can} describe such a landscape.

 The solution is associated to a matter relevant vertex operator
$\phi_u(z)$. Here $u$ parametrizes the RG flow obtained by
integrating $\phi_u$  on the boundary of the unit disk,
\be
S^{(u)}=S_0+\int_{\del Disk} d\theta \phi_u(\theta),
\ee
 with $S_0$ corresponding to the original BCFT and with the conventions that
  $u=0$ corresponds to the UV (that is $\phi_{u=0}=0$) and $u=\infty$ to the IR,
  which is possible for all the explicit cases we consider.\\
In order to avoid the subtle issue of renormalization of UV
divergences (which occur for general tachyon profiles) we will
consider the case in which the relevant field $\phi_u$ is subject to
the condition \be (c\phi_u)^2=0\label{1}, \ee where $c$ is the
reparametrization ghost. If  $\phi_u$ were a marginal field, the
above condition would rule out any non trivial OPE of $\phi_u$ with
itself. In our case, however, the field $\phi$ can  have divergent
OPE and still satisfy (\ref{1}). This is because if $\phi$ is
relevant enough, its contact term divergence can be weaker than the
vanishing of the $cc$ collision. In the case where $\phi$ is made of
primary operators
$$\phi=\sum_i \chi^{(h_i)}_i,$$
 this is realized whenever all of them have weights
 $$h_i<\frac12,$$
 see \cite{Fendley} for an explicit example.
The region $\frac12\leq h<1$ should be explored with a consistent
 renormalization method as in the case of marginal deformations, \cite{Kiermaier:2007vu}.
  We do not try to attack this problem in the present paper.

The solution can be written within a minimal extension of the
 universal $K,B,c$ algebra, \cite{Okawa1, Erler:2006hw, Erler:2006ww} where
 we also add the identity--based
insertion (defined in the $\frac2\pi\arctan$--sliver frame)
$$\phi_u=\phi_u\left(\frac12\right)I,$$
where $I$ is the identity string field. It takes the form
\be \psi_u=c\phi_u-\frac B{K+\phi_u}\,u\del_u\phi_uc\del c. \ee The
$u$ dependence turns out to be a gauge artifact as
$u$ can be changed by a simple scale transformation in the sliver
frame (which is a midpoint preserving reparametrization).  The
 structure is  similar to a reparametrization of the  Erler--Schnabl Tachyon Vacuum ($TV$),
\cite{ErlerSchnabl}
\be \psi^u_{TV}=\frac u{u+K}c(u+K)Bc=u c-\frac u
{u+K}Bc\del c,\quad u>0,
\ee
which, for reparametrization parameter $u=1$, reduces to (the `left' version of)  \cite{ErlerSchnabl}.
In particular the tachyon vacuum
corresponds to the simplest choice of $\phi_u$
\be \phi_u^{(TV)}=u.
\ee

 We show that if the string field $\frac1{K+\phi_u}$ is regular,
then the solution still represents the tachyon vacuum, although in a
non--universal form. So non trivial RG fixed points are described
when $\frac1{K+\phi_u}$ is {\it singular} but
$\frac B{K+\phi_u}\,u\del_u\phi_u$ is {\it finite} (since the
solution itself must be finite). Remarkably, the trace of the matter part of this last
quantity turns out to coincide with the shift in the partition
function from the UV to the IR on the canonical cylinder of width 1.
\be
Tr[\frac1{K+\phi_u}\,u\del_u\phi_u]=(\lim_{u\to0}-\lim_{u\to\infty})\left\langle
e^{-\int_0^1ds\,\phi_u(s)}\right\rangle_{C_1}. \ee Thanks to this
important identity, the closed string overlap, \cite{Ellwood:2008jh}, is shown to coincide
with the shift in the disk--one--point function of an on--shell
closed string.  Similar considerations apply for the on--shell value
of the action. We show that the energy is independent of the gauge
parameter $u$ and we write down a closed regular expression for it.
Due to the essentially more involved structure wrt the closed string
overlap, we have postponed its explicit calculation.

 The paper is organized as follows. In section 2 we show that all we are going to
  discuss is already hinted at by  the universal tachyon vacuum solutions, in a very simple way:
  this is expected since the tachyon vacuum is the simplest IR fixed point that one gets by integrating
  on the boundary of the worldsheet a zero momentum tachyon. We take the time to make some simple yet
  important observation about the  Erler--Schnabl solution, \cite{ErlerSchnabl}. Thanks to the structure of the solution,
  it is very easy to identify the world--sheet RG flow and to naturally associate the UV fixed point with
   the identity string field and the IR with the sliver, where a zero momentum tachyon is integrated along
   the boundary. We show how the integration over the Schwinger parameter, upon evaluation of gauge
    invariant quantities, reduces to the shift between the UV and IR of the corresponding observables.

In section 3 we algebraically build the solution, by writing a pure
gauge ansatz, inspired by the reparametrizations of the
Erler--Schnabl tachyon vacuum. To do so we minimally extend the
$K,B,c$ algebra to contain the identity--based insertion of a
relevant operator and its BRST transform. No further extension is
needed to close the multiplicative differential algebra. Inside this
algebra we point out how to build wedge--like surfaces, where the
relevant operator is exp--integrated along the boundary. The
solution we write down is somehow similar to the Erler--Schnabl
tachyon vacuum in the sense that it is a continuous superposition of
wedge states, however such wedges have  $deformed$ boundary
conditions
 given by exp--integrating the relevant field on the boundary. Given this crucial difference wrt
 universal tachyon vacuum solutions, we nonetheless show that the  solution  is  still `very close'
  to just be  the same tachyon vacuum, expressed in a non--universal way. We then state the crucial
  conditions under which we can have a  solution which is different from the tachyon vacuum.

In section 4 we determine some sufficient conditions which allow to
relate the above--mentioned  conditions to the  IR limit
of the partition function along the RG flow: the disk partition
function should be finite and not vanishing in the IR. As a matter
of fact, this crucial requirement has been already pointed out by
Ellwood, \cite{Ellwood:2009zf}, in order to guarantee the existence
of a finite characteristic projector.  We then show how the shift in
the partition function is encoded inside the solution. A consequence
of this, section 5, is the fact that the Ellwood invariants of the
solution correctly capture the shift in the closed string one--point
function between the UV BCFT and the IR one. In section 6 we write
down the energy functional of the solution, and we
 discuss it to some extent.

In section 7 we give concrete examples of solutions describing well
known RG-flows: lower dimensional
branes in non compact and compact dimensions. We end up with some conclusions
 and open problems.

A few appendices contain additional material: the construction of
the solution using the homotopy field at the tachyon vacuum
(Appendix A), the explicit proof that
 the two non real forms of the solution have the same energy and that the
 energy can indeed be
 written as the integration of an explicitly real string field
 (Appendix B), the generalization of the solution to
 the (cubic) superstring case (Appendix C, with the left unanswered interesting
 question on wheatear
 cubic superstring field theory  contains  solutions which are not expected from string theory,
 such as codimension 1 D--branes in the GSO(+) sector). Some explicit computations for the simplest
 relevant deformation based on $\int:X^2:$ are reviewed and presented in Appendix D.
 Finally, Appendix E is the derivation of equation \eqref{orderedEXP}.\\

\section{Reparametrizations of the Erler--Schnabl Tachyon Vacuum}

The quickest way to arrive at the `relevant' solutions is a very simple yet illuminating observation about the recent solution derived by Erler and Schnabl, describing the tachyon vacuum. The solution, \cite{ErlerSchnabl}, is given by
\be
\psi_0=\frac1{1+K}c(1+K)Bc=c-\frac1{1+K}Bc\del c,
\ee
and can be formally  obtained with a singular gauge transformation of the perturbative vacuum
\be
\psi_0&=&U_0QU_0^{-1}\\
U_0&=&1-\frac1{1+K}Bc\\
U_0^{-1}&=&1+\frac1{K}Bc.
\ee
A  trivial infinite class of gauge equivalent solutions can be obtained by a simple scale transformation
(which, on the cylinder, is a
midpoint preserving reparametrization)
\be
z\to \frac zu
\ee
under which the $KBc$ algebra is mapped to an isomorphic (i.e. equivalent) representation
\be
c&\to&u c\\
(B,K)&\to&\frac1u (B,K).
\ee
The new solutions are thus generated by the gauge transformations
\be
U_u&=&1-\frac u{u+K}Bc\\
U_u^{-1}&=&1+\frac u{K}Bc,
\ee
 giving the reparametrizations  of the tachyon vacuum
\be
\psi_u=\frac u{u+K}c(u+K)Bc=u c -\frac u{u+K}Bc\del c.
\ee
Since reparametrizations do not change the bpz norm (and consequently the gauge invariant action and closed string overlap) we have
\be
\langle\psi^{(u)},Q\psi^{(u)}\rangle&=&\langle\psi^{(u=1)},Q\psi^{(u=1)}\rangle\0\\
\langle V_c,\psi^{(u)}\rangle&=&\langle V_c,\psi^{(u=1)}\rangle, \ee
as can also be explicitly checked by rescaling the involved
correlators on the cylinders.

A naive extrapolation would erroneously lead us to believe  the
above result to hold for $any$ value of $ u$. However
 we must set  $u>0$ in order to  ensure that, under the scaling, all wedge
 states ($e^{-t K}\to e^{-\frac tu K}$)  remain of finite
positive width. A negative value of $u$ would result in inverse
wedge states (wedge states with negative width), which are
ill--defined objects ($e^{tK}$). One must also note that blindly
setting $u=0$ {\it is not} the same as taking the limit $u\to0^+$.
In the former case the gauge transformations become the identity and
the perturbative vacuum (D25 brane) remains trivially untouched; but
by taking the $u\to0^+$ limit a phantom piece is switched on \be
\lim_{u\to0^+}\frac u{u+K}=sliver \ee and the corresponding solution
is still the tachyon vacuum (although singularly
 reparametrized).
To summarize, we have \be
u>0\, &\rightarrow&\, \textrm{Tachyon Vacuum}\0\\
u\equiv0\, &\rightarrow&\, \textrm{D25--brane}\0\\
u<0\, &\rightarrow&\, \textrm{Ill--defined solution}.\0
\ee

The above observation has a simple physical interpretation in terms
of the tachyon potential. The gauge parameter $u$ is like an initial
displacement applied to the system at the perturbative vacuum in
order to reach another stationary point. For $u>0$ the displacement
is in the right direction to reach the tachyon vacuum. For $u=0$
there is no displacement at all, so the solution will stay at the
perturbative vacuum. For $u<0$ the displacement is in the wrong
direction, and the solution falls in the unbounded region of the
potential, becoming singular.

This target-space perspective can be brought (and made more precise) to the worldsheet. The trivial gauge parameter $u$ can be interpreted as the
value during the RG flow of a zero momentum tachyon integrated along the boundary. To see this, the crucial quantity to analyze is
\footnote{A BRST invariant insertion of $Y(z)=\frac12\del^2 c\del c c(z)$ is understood to saturate the ghost number. The reader can easily verify that this simple trace captures all the essential features of the closed string overlap, in particular the shift from the UV to the IR.}
\be
Tr[\frac u{u+K}]=\int_0^{\infty}dTu\left\langle  e^{-u T}\right\rangle_{C_{T}}=\int_0^{\infty}dT\;u\left\langle  e^{-\int_0^Tds\, u}\right\rangle_{C_{T}},
\ee
rescaling to a canonical cylinder of width 1
\be
Tr[\frac u{u+K}]&=&\int_0^{\infty}dT\;u\left\langle  e^{-\int_0^1ds\, (Tu)}\right\rangle_{C_{1}}\0\\
&=&\int_0^{\infty}dT (-\del_T)\left\langle  e^{-\int_0^1ds\, (Tu)}\right\rangle_{C_{1}}\0\\
&=&\left(\lim_{u\to0^+}-\lim_{u\to\infty}\right)\left\langle\,e^{-\int_0^{2\pi}\frac{d\theta}{2\pi}\, u}\right\rangle_{Disk}=Z_{u=0}-Z_{u=\infty},
\ee
where by
\be
Z(u)\equiv\left\langle\,e^{-\int_0^{2\pi}\frac{d\theta}{2\pi}\, u}\right\rangle_{Disk}=e^{-u},
\ee
we denote the disk partition function obtained by integrating a zero momentum tachyon on the disk (or cylinder) boundary.\\
We see that the quantity we have computed is just the shift in the
partition function between the perturbative vacuum (UV fixed point)
and the tachyon vacuum (IR fixed point), where the partition
function identically vanishes due to the infinite suppression given
by the zero momentum tachyon which is sent to $\infty$ during the RG
flow.

 Needless to say  this is  the way the tachyon condenses in
Boundary String Field Theory, but the reader should not be misled by
this: in BSFT the zero momentum tachyon field extremizes the action
only when $u=0$ (no deformation present) or when $u=\infty$
(endpoint of the RG flow). In cubic OSFT --more precisely, in the
Erler--Schnabl description of the tachyon vacuum-- $u$ is a trivial
gauge parameter: the shape of the tachyon field is meaningless (it
is not gauge invariant) but the gauge invariant observables exactly
capture the entire jump between the UV and the IR of the worldsheet
RG flow, independently of the initial value of $u$. In other words,
the Erler-Schnabl solution (with its reparametrizations) is just
what is needed to
explicitly `see' the RG flow at work on the world-sheet and, in this sense,
 it can be considered a `BSFT'-like solution. It would be interesting to
 generalize this discussion to other universal solutions (for example to the
 original Schnabl solution, where a simple reparametrization seems to give
 much less interesting structure).

Another important remark is about  the role of the star algebra projectors in this mechanism: the
integration over the Schwinger parameter
reduces to the difference between the norm of the identity string field (which gives the UV fixed point)
and that of the sliver with a relevant boundary integration (which gives the IR fixed point). It is really
the sliver limit that drives the RG flow all the way to the IR. This is yet another nice property
of sliver--like projectors inside the star algebra: rescaling to finite width a very large wedge state,
 dressed with a relevant boundary exp-integration,  brings the relevant deformation to its IR fixed point,
 in the sliver limit.\footnote{ In fact this was already noted by Rastelli, Sen and Zwiebach,
 \cite{Rastelli:2001vb}, during the `VSFT' era. A consequence of this is that the deformed sliver will
  have the boundary conditions of the IR BCFT at its midpoint, \cite{Ellwood:2009zf}.}

These  considerations suggest  that a lump solution can
be obtained by replacing the gauge parameter $u$ with a properly
chosen tachyon profile $\phi(x)$. Heuristically, from a target space
perspective, $\phi(x)=0$ will correspond to regions in the
transverse space where the theory is  at the perturbative vacuum,
while for $\phi(x)>0$ the theory will be at the tachyon vacuum. More
precisely, in the far infrared of the worldsheet, the minima of
$\phi(x)$ will correspond to the regions where all the energy will
be stored, \cite{Harvey}, that is the location of the lower
dimensional branes. However, since $X$ is a quantum field and not a
number, the height of the minima of $\phi(x)$ should be very
carefully balanced in order to have a well defined (and non trivial)
IR fixed point.  We also learn that a `localized' lump solution
(i.e. a  tachyon with a localized dependence on the $X$--zero mode)
is not a  requirement as, already for the tachyon vacuum, the
tachyon VEV can be freely changed (and even pushed to infinity) by a
scale transformation in the arctan frame.

\section{The solution: general structure}
\subsection{Adding a relevant field to the $K,B,c$ algebra}

The solution lives in a minimal extension of the universal $K,B,c$ subalgebra, which can be defined in the sliver frame (obtained by mapping the UHP to an infinite cylinder $C_2$ of circumference 2, by the sliver map $f(z)=\frac2\pi\arctan z$) as\footnote{Local insertions of fields are defined on $C_2$ so that, for primaries, $\chi_h^{(C_2)}(1/2)=\pi^h \chi_h^{(UHP)}(1)$.}
\be
K&=&\frac\pi2K_1^L\ket I\\
B&=&\frac\pi2B_1^L\ket I \\
c&=& c\left(\frac12\right)\ket I.
\ee
They obey the well known star--product relations
\be
\{B,c\}&=&1\\
\,[K,B]&=&0\\
\,[K,c]&\equiv&\del c\\
\{B,\del c\}&=&0,
\ee
as well as the differential relations
\be
QB&=&K\\
Qc&=&cKc=c\del c.
\ee
A relevant {\it matter} operator $\phi$ (not necessarily primary) is now added.
As for $K,B,c$ we introduce it as an identity--based insertion in the sliver frame.
\be
\phi=\phi\left(\frac12\right)\ket I.
\ee
The new simple $*$--product relations  are given by
\be
\,[c,\phi]&=&0\\
\,[B,\phi]&=&0\\
\,[K,\phi]&\equiv& \del\phi,
\ee
as well as
\be
\,[c,[K,\phi]]&=&[c,\partial\phi]=0\\
\,[\phi,[K,c]]&=&[\phi,\partial c]=0\\
\,[B,[K,\phi]]&=&[B,\del\phi]=0. \ee The BRST charge acts on $\phi$
in the following way \be Q\phi=c\del\phi+\del c\delta\phi. \ee When
$\phi$ is a primary of weight $h$ we simply have \be
Q\phi=c\del\phi+h \del c\phi. \ee If $\phi$ is not primary, then
$\phi$ needs not be proportional to $\delta\phi$, which can be
explicitly computed from the BRST variation. In order to avoid the
explicit appearance of $\del\phi$ in the solution we will write down
below, it is useful to introduce the insertion $\phi$ together with
a $c$ ghost in front. This gives \be Q(c\phi)=c\del c(\phi-\delta
\phi). \ee If $\phi$ is a primary of dimension $h$ we then have \be
Q(c\phi)=c\del c(1-h)\phi, \ee which correctly vanishes when $\phi$
is marginal.

An important assumption we will make about the matter insertion $\phi$ is that its OPE should be such as to make\footnote{The assumption is actually  more general. In order to define deformed wedge states without the need of a renormalization of contact term divergences, see later, we need the non-local operator $e^{-\int_a^bds\, \phi(s)}$ to be finite by itself. This  (at least in conformal perturbation theory) is an assumption about the finiteness  of  the integrated $N$--point functions $\int ds_1...\int ds_N\langle \phi(s_1)...\phi(s_N)\rangle^{(BCFT_0)}$, which is satisfied in all the explicit examples we consider in the present paper. }
\be
(c\phi)^2=(c\phi)(c\delta\phi)=(c\delta\phi)(c\phi)=0\label{cphi}
\ee
In the case where $\phi=\phi^{(h)}$ is a boundary primary field of weight $h$, with a singular OPE on the UHP of the kind
\be
\phi(x)^{(h)}\phi(0)^{(h)}\sim \frac a{|x|^{2h}},\quad x\in\mathfrak{R}\label{ope}
\ee
 this implies that
\be h_\phi<\frac12. \ee In particular, differently from the marginal
case, here (mild) singular collisions between the $\phi$'s are
allowed.\footnote{It is clear that one can also consider  linear combinations of primaries with global regular OPE, with the only restriction that they should be relevant ($0\leq h<1$), this could be useful, for example, to get non--marginal time dependent solutions.}

Inside this algebra one can build wedge--like surfaces, where the
matter field $\phi$ is integrated along the boundary. To do this one
simply adds $\phi$ to the strip--generator $K$
\be K\to K+\phi,
\ee
and exponentiate it, see also \cite{simple-marg}. That is, a deformed wedge state of width
$T$ is  given by
\be \ket {T,\phi}=e^{-T(K+\phi)}.
\ee
 The overlap
with a Fock space test state $\chi=\chi(0)\ket0$ is given by
\be
\bra\chi\ket{T,\phi}=\left\langle f\circ \chi(0)
e^{-\int_{\frac12}^{T+\frac12}ds\, \phi(s)}\right\rangle_{C_{T+1}},
\ee
 where
 $$f(z)=\frac2\pi\arctan z,$$
 and we have used the fact
that, inside the path integral, we can represent (see Appendix E or \cite{simple-marg} for
a derivation)
\be
\langle[...]e^{-T(K+\phi)}[...]\rangle=\langle[...]e^{-TK}e^{-\int_x^{x+T}ds
\phi(s)}[...]\rangle. \label{orderedEXP}
\ee
In other words,
just as $e^{-TK}$ generates a strip of worldsheet with unmodified
boundary conditions, $e^{-T(K+\phi)}$ generates the same strip but
it also inserts in the path integral $e^{-\int_x^{x+T}ds \phi(s)}$,
which gives modified (non conformal unless $\phi$ is exactly
marginal) boundary conditions. This operation requires a
regularization and a consequent renormalization in the case of
marginal $\phi$ with singular OPE, but it is well defined  when
$\phi$ is relevant enough, i.e. it satisfies (\ref{cphi}), see \cite{Fendley} for an explicit
example of how the divergences in the $N$--point functions of $\phi$ gives rise to integrable singularities, with an overall finite result. We do not attempt here to find a regularization procedure
to treat all kind of relevant fields, but we will stick to
(\ref{cphi}).

It is important to be aware that, unless $\phi$ is BRST closed and
(since it is a matter field) proportional to the identity operator
(that is, a pure number) the deformed wedge state is not BRST closed
\be Q \ket {T,\phi}=Q e^{-T(K+\phi)}=\int_0^Tdt
e^{-t(K+\phi)}(-Q\phi)e^{-(T-t)(K+\phi)}\neq 0. \ee This is also
true when $\phi=j$ is marginal, however in that case the above
integral can be recast in \be Q(e^{-T(K+j)})=-[cj,e^{-T(K+j)}], \ee
which implies that the BRST variation of the deformed wedge is just
a `boundary' contribution: this is indeed what happens by deforming
the boundary conditions while preserving conformal invariance,
\cite{Kiermaier:2007ki}. In the case $\phi$ is relevant, on the
other hand, the BRST violation has also a term which cannot be
pushed to the extremes of the wedge \be
Q(e^{-T(K+\phi)})=-[c\phi,e^{-T(K+\phi)}]+\int_0^Tdt\,e^{-t(K+\phi)}\,[K,c](\phi-\delta\phi)\,e^{-(T-t)(K+\phi)}.
\ee This is the reflection of the fact that the boundary conditions
given by inserting $e^{-\int\phi}$ in the path integral are not
conformal.\footnote{In the trivial  case where $\phi$ is a non zero
constant the two pieces in the rhs of the above equation identically
cancel, however this does not mean that the boundary conditions are
conformal.}

 The possibility of building wedge states with deformed
(non conformal) boundary conditions is the basic property thanks to
which a boundary RG flow can be engineered inside a minimal
extension of the $K,B,c$ algebra: by gluing together the deformed
wedge states, one generates surfaces where a relevant operator is
integrated along the boundary. The original scale invariance of the
world sheet path integral will now generate an RG flow which will
drive the system from the perturbative vacuum, where our OSFT is
defined, to a new conformal fixed point.

\subsection{Solution from pure gauge ansatz}

Two simple non real solutions can be constructed, which are the `reality'--conjugate of one another (they can be obtained from one another by reading them in the
opposite sense). Focusing on the `left' solutions (as in the previous section) we write down the following gauge transformation
\be
U_\phi&=&1-\frac1{K+\phi}\phi Bc\label{gaugetransf}\\
U_\phi^{-1}&=&1+\frac1{K}\phi Bc, \ee the  collisions between the
$\phi$'s in the denominator and in the numerator of $U$ do not give
rise to divergences if (\ref{cphi}) holds, i.e. if
$\phi(z)\phi(0)=o(|z|^{-1})$.

Notice that these gauge transformations are obtained by substituting
the $u$ of the previous section with the relevant field $\phi$. In
doing this there are ambiguities in the position of $\phi$ and $K$
(the two operators do not commute). In this particular order both
$U$ and $U^{-1}$ are relatively simple, and they do not give rise to
$1/K$ factors in the solution. In Appendix A we show how these gauge
transformations can be obtained from a singular gauge transformation
of the tachyon vacuum, following the procedure suggested by Ellwood, \cite{Ellwood:2009zf}.
Here we just stick to this form, leaving open the interesting
question about the existence of other kinds of gauge transformations
that could result in more general solutions. We call the
relevant field $\phi$, to which the solution is associated, the
$seed$.

More importantly, notice that, because of the explicit $1/K$,
$U^{-1}$ is certainly singular (except for the trivial case
$\phi\equiv0$), while $U$ can be regular or singular according to
the behavior of the string field $\frac1{K+\phi}$. As we will show
in a while, if this string field turns out to be regular, then the
solution will just be a non universal way of writing down the
tachyon vacuum. In fact, see next subsection, only a singular
$\frac1{K+\phi}$ can give rise to a new solution. Anticipating the
results of the next section, this is related to the fact that the IR
partition function of the relevant boundary deformation given by the
integration of $\phi$ should be finite and not vanishing in the IR.
A vanishing partition function in the IR would mean that the IR
fixed point that has been reached is still the tachyon vacuum and
not a lower dimensional brane (or a new genuinely propagating BCFT).

Using the BRST variation \be
Q(c\phi)=(\phi-\delta\phi)c[K,c]=(\phi-\delta\phi)c\del c\0 \ee we get
\be
\psi_{\phi}=U_\phi Q U_\phi^{-1}&=&\left(1-\frac1{K+\phi}\phi Bc\right)Q\left(1+\frac1{K}\phi Bc\right)\0\\
                                &=&\left(1-\frac1{K+\phi}\phi Bc\right)\left(c\phi-\frac B{K}(\phi-\delta\phi)c\del c\right)\0\\
&=&c\phi-\frac1{K+\phi}(K+\phi-\phi)\frac1K(\phi-\delta\phi) Bc\del c\0\\
&=&c\phi-\frac1{K+\phi}(\phi-\delta\phi) Bc\del c.
\ee
Notice that there is a naked identity--based piece in the solution.
However, because  $c\phi$ has negative net scaling dimension (since $\phi$ is
relevant), the typical ambiguities related to identity--based string fields
are here avoided. \footnote{As explained in \cite{ErlerSchnabl}, this can be
 made more precise by understanding the seed $\phi$ as $$\phi=\lim_{\epsilon
 \to0}\phi\Omega^\epsilon, \quad\quad \Omega= e^{-K}.$$  }

It is important to observe that, in the case where $\phi$ is a
marginal operator (with trivial OPE, so that no renormalization of
the boundary integral is needed) only the identity piece $c\phi$
survive ($\phi-\delta\phi=0$). This is a trivial solution of the
equation of motion but, due to its vanishing scaling dimension,
the associated observables (energy and closed string overlap) are
undefined (they depend on the way the limit $\epsilon\to0$ is
taken). So, from the point of view of identity-like singularities, our solution is only acceptable for a truly relevant $\phi$.
Remember also that this solution  describes a universal tachyon
vacuum by simply choosing a constant, positive $\phi$ (the zero
momentum tachyon).

As anticipated two (non real) solutions can be constructed
\be
U_L&=&1-\frac1{K+\phi}\phi Bc\\
\psi_L&=&U_LQU_L^{-1}=c\phi-\frac1{K+\phi}(\phi-\delta\phi)Bc\del c\label{tpsiL}\\
.\0\\
U_R&=&1- cB\phi\frac1{K+\phi}\\
\psi_R&=&U_R^{-1}QU_R=c\phi -c\del cB(\phi-\delta\phi)\frac1{K+\phi}=\psi_L^\ddag\label{tpsiR},
\ee
where $\ddag$ is the reality--conjugation, \cite{Gaberdiel:1997ia}
\be
\ddag\equiv \textrm{bpz}^{-1}\circ\textrm{hc}.
\ee

Since the two solutions are written in pure gauge form, they will, at least formally, automatically solve the equation of motion. With some manipulations in the
$K,B,c;\,\phi$ algebra one can show that\footnote{Here we are {\it assuming} that it is possible to define the string field $\left[\frac1{K+\phi}(\phi-\delta\phi)\right]$ such that $$(K+\phi)\left[\frac1{K+\phi}(\phi-\delta\phi)\right]=\phi-\delta\phi.$$
See \cite{EM} and \cite{BGT, BGT3}, where this problem is discussed, using two different attitudes.}
\be
Q\psi_L&=&\left(c\phi -\frac1{K+\phi}(\phi-\delta\phi)\del c\right)\frac1{K+\phi}(\phi-\delta\phi)Bc\del c=-\psi_L^2
\ee
and accordingly
\be
Q\psi_R&=&c\del cB(\phi-\delta\phi)\frac1{K+\phi}\left[c\phi+\del c(\phi-\delta\phi)\frac1{K+\phi}\right]\0\\&=&
c\del cB(\phi-\delta\phi)\frac1{K+\phi}\psi_R=-\psi_R^2\label{QpsiR}.
\ee
Notice that, differently from the tachyon vacuum solution, two Schwinger
parameters are needed for $Q\psi$. Due to the particular form of the
solutions,
and to the commutation relations of the $K,B,c,\phi$ algebra
\be
\{B,\del c\}&=&0\\
\,[B,K]=[B,\phi]&=&[c,\phi]=[\del c,\phi]=0,
\ee
it is easy to prove that the two solutions share the same energy
\be
\textrm{tr}[\psi_R\psi_R\psi_R]=\textrm{tr}[\psi_L\psi_L\psi_L],
\ee
and the same closed string overlap
\be
\textrm{tr}[V_c\psi_R]=\textrm{tr}[V_c\psi_L].
\ee
In Appendix B an explicit real form of the energy is given.
With some algebra one can also check that the two solution are related by a very simple gauge transformation in the $K,B,c,\phi$ algebra
\be
\psi_L=\frac1{K+\phi}(Q+\psi_R)(K+\phi).
\ee
This allows to write (quite formally) a real solution as
\be
\psi=\psi^\ddag=\frac1{\sqrt{K+\phi}}(Q+\psi_R)\sqrt{K+\phi}=\sqrt{K+\phi}(Q+\psi_L)\frac1{\sqrt{K+\phi}}.
\ee

\subsection{Is this different from the Tachyon Vacuum?}

It is instructive to write down the kinetic operator around this
solution  to see if it is possible to trivialize it. With some
manipulation, using the $K,B,c,\phi$ algebra it is possible to show
that
\be {\cal Q}_{\psi_\phi} \frac B{K+\phi}=Q\frac
B{K+\phi}+\left\{\psi_\phi,\frac B{K+\phi}\right\}=1.
\ee
So, unless
the homotopy--field $\frac B{K+\phi}$ is ill-defined (as it is the case
for $\frac BK$ and the original $Q$), the solution has trivial
cohomology, which is the defining property of the tachyon vacuum
\cite{Ellwood}. On the other hand, in order for the solution to be
finite, the quantity $\frac B{K+\phi}(\phi-\delta\phi)$ should
be itself finite (at least in the `weak' sense that it should give rise to well
defined observables, and finite overlaps with Fock-space states). In full generality we thus have a new
nontrivial solution if
\begin{itemize}
\item $\frac1{K+\phi}$ is divergent
\item $\frac B{K+\phi}(\phi-\delta\phi)$ is finite.
\end{itemize}
We will now give some sufficient conditions for the seed $\phi$ in order to
comply with the above requirement.

\section{A class of nontrivial  solutions}

We introduce here a general class of seeds $\phi$ that give rise to Fock-space-finite
non trivial solutions. We will later give explicit examples of
seeds in this class. Let $u$ be a parametrization of the
world--sheet RG flow represented by integrating a relevant vertex
operator $\phi_u$ on the boundary. We take $u$ to span the positive
real axis with $u=0$ corresponding to the UV (no deformation
present) and $u=\infty$ to the IR. In particular we have \be
\phi_{u=0}=0.\label{cnd0} \ee
The first important condition is that
$u$ should linearly increase under scaling in the cylinder
coordinate frame. That is \be f_t(z)=\frac zt\0 \ee should imply \be
f_t\circ\phi_u(z)=\frac1t\,\phi_{tu}\left(\frac
zt\right).\label{cnd1} \ee
This condition actually states that $u$ can be changed to any positive value by a scale transformation. Thus $u$-dependence of the solution will be a gauge artifact.\\
The second, related, condition is about the BRST variation of
$c\phi_u$. We will require that \be Q(c\phi_u)=c\del
c(\phi_u-\delta\phi_u)=c\del c\,u\del_u\phi_u.\label{cnd2} \ee
Notice that the above conditions are trivially satisfied for the
seed giving rise to the tachyon vacuum, where one just chooses \be
\phi_u=u>0. \ee The third condition is related to our aim to avoid
flowing to the tachyon vacuum in the IR, that is to the requirement
that at the IR fixed point some perturbative open string dynamics is
left. In order for this to happen the partition function on the
canonical cylinder of width 1 (or equivalently on the unit disk)
should be finite and not vanishing, at the end of the RG-flow \be
\lim_{u\to\infty}\left\langle e^{-\int_0^1ds\, \phi_u(s)
}\right\rangle_{C_1}=\verb"finite".\label{cnd3} \ee The importance
of this condition, which is usually ignored in the literature on
world--sheet RG flows, has been originally stressed by Ellwood,
\cite{Ellwood:2009zf}, whose intuition is here made concrete.

Now we want to show that these conditions are just enough  to guarantee that $\frac1{K+\phi_u}$ is $ \textrm{divergent}$ while $Tr[\frac1{K+\phi_u}(\phi_u-\delta\phi_u)]$ is $ \textrm{well defined}.$
Let us first consider
\be
Tr[\frac1{K+\phi_u}]=\int_0^\infty dT\,\left\langle e^{-\int_0^Tds\, \phi_u(s) }\right\rangle_{C_T}=(*),
\ee
by a scale transformation, using (\ref{cnd1}) we get
\be
(*)= \int_0^\infty dT\,\left\langle e^{-\int_0^1ds\, \phi_{Tu}(s) }\right\rangle_{C_1}=\int_0^\infty dT\, g(Tu),
\ee
where we denoted
\be
g(v)\equiv Tr[e^{-(K+\phi_v)}]=\left\langle e^{-\int_0^1ds\, \phi_{v}(s) }\right\rangle_{C_1},
\ee
the partition function on the canonical cylinder obtained by integrating $\phi_v$ on the boundary.\\
This partition function is in turn identified with the exp of boundary entropy  and from the $g$--theorem/conjecture, \cite{Affleck},  we know that $g(v)$ is a monotonically decreasing (and positive) function of $v$. But because of condition (\ref{cnd3}) $g(uT)$ will asymptote to a finite non vanishing positive value for $T\to\infty$. Thus the integral diverges. This shows that the string field $\frac1{K+\phi_u}$ is singular as required for the solution to be different from the tachyon vacuum.\\
Let us now analyze the trace of the quantity entering the solution, that is
\be
Tr[\frac1{K+\phi_u}(\phi_u-\delta\phi_u)]=Tr[\frac1{K+\phi_u}u\del_u\phi_u]=\int_0^\infty dT\,Tr[ e^{-T(K+\phi_u)}u\del_u\phi_u],
\ee
where condition (\ref{cnd2}) has been used.
Recalling the derivative of a non--commutative exponential
\be
d(e^M)=\int_0^1ds\, e^{sM}\,(dM)\,e^{(1-s)M},\0
\ee
and simplifying it inside a  trace
\be
Tr[d(e^M)]=Tr[\int_0^1ds\, e^{sM}\,(dM)\,e^{(1-s)M}]=\int_0^1dsTr[(dM)e^{(1-s+s)M}]= Tr[(dM)e^{M}],\0
\ee
we have
\be
u\int_0^\infty dT\,Tr[\del_u\phi_ue^{-t(K+\phi_u)}]&=&-\int_0^\infty dt\,\frac ut \del_u Tr[e^{-t(K+\phi_u)}]\0\\
&=&-\int_0^\infty dt\,\frac ut \del_u \left\langle e^{-\int_0^T\,\phi_u(s) ds}\right\rangle_{C_T}\0\\
&=&-\int_0^\infty dt\,\frac ut \del_u\,g(tu)=-\int_0^\infty dx\,
\del_x\,g(x). \ee

To summarize, we have the very remarkable property \be
Tr[\frac1{K+\phi_u}(\phi_u-\delta\phi_u)]&=&g(0)-g(\infty)=Z^{UV}-Z^{IR}.
\ee Not only is this quantity perfectly well defined but it also
precisely agrees with the shift in the open string partition
function from the UV to the IR!\footnote{If we work in flat non
compact space, then $g(0)$ will obviously diverge because of the
infinite volume, we do not consider this to be a sign of `non
regularity', since it is just the reflection that $X$ has a non
compact zero mode.} Notice that we have  used the conditions
(\ref{cnd1},\ref{cnd2},\ref{cnd3}) to arrive at this remarkable
result. Whether these conditions are also necessary, or can be
modified in some way, is an interesting problem we leave for further
investigation. It turns out, see \cite{EM,BGT}, that finiteness of
$Tr[\frac1{K+\phi_u}(\phi_u-\delta\phi_u)]$ is enough to guarantee
finiteness of  $\psi_L$ inside observables and against Fock space states.

\section{Closed String overlap}

The simplest gauge invariant quantity that can be explicitly computed for the class of solutions defined in the previous section is the closed string overlap, i.e.
the overlap between the solution and an on--shell closed string state inserted at the midpoint of the identity string field. According to Ellwood, \cite{Ellwood:2008jh}, this quantity should be equal to the shift
in the closed string one--point function between the new BCFT represented by the new solution and the reference BCFT represented by the
perturbative vacuum. In full generality, if $\psi_1$ represents the new $BCFT_1$, expressing everything on a canonical cylinder of
width 1, we expect to find
\be
Tr[V_c\,\psi_1]=\left\langle V_c(i\infty)c(0)\right\rangle^{(BCFT_0)}_{C_1}-\left\langle V_c(i\infty)c(0)\right\rangle^{(BCFT_1)}_{C_1}.
\ee
In the case  $BCFT_1$ is the IR fixed point of a relevant boundary interaction, the correlator in $BCFT_1$ can be seen as a correlator in $BCFT_0$ {\it with an exp--boundary integral}, responsible for the change in
boundary conditions from  $BCFT_0$ to $BCFT_1$.
\be
\left\langle V_c(i\infty)c(0)\right\rangle^{(BCFT_1)}_{C_1}=\lim_{v\to v^*}\left\langle V_c(i\infty)c(0)\,e^{-v\int_0^1dt\,\theta(t)}\right\rangle^{(BCFT_0)}_{C_1},
\ee
where $v^*$ is the IR fixed point in coupling space.
\no Therefore
\be
Tr[V_c\,\psi_1]&=&\left\langle V_c(i\infty)c(0)\,\left(1-\lim_{v\to v^*}e^{-v\int_0^1dt\,\theta(t)}\right)\right\rangle^{(BCFT_0)}_{C_1}\0\\
&=&Tr\left[V_c\,c\left(\Omega-\lim_{v\to v^*}\tilde\Omega_v\right)\right],
\ee
where we have used the {\it boundary--changed} vacuum $\tilde\Omega_v$
\be
\Omega&=&e^{-K}\\
\tilde\Omega_v&=&e^{-K-v\theta}.
\ee

Our solution is given by \be
\psi_u=c\phi_u-\frac1{K+\phi_u}(\phi_u-\delta\phi_u)Bc\del
c=c\phi_u-\frac1{K+\phi_u}u\del_u\phi_uBc\del c, \ee where we have
already chosen a seed $\phi_u$ such that (\ref{cnd2}) is satisfied.
The identity piece in the solution does not contribute, thanks to
the globally negative scaling dimension (the seed is relevant). In
particular, using the scaling property (\ref{cnd1}), we have \be
Tr[V_c\,c\phi_u]=\lim_{\epsilon\to0}Tr[V_c\,c\phi_u\Omega^\epsilon]=\lim_{\epsilon\to0}Tr[V_c\,c\phi_{\epsilon
u}\Omega^1]=0, \ee because \be \phi_{u=0}\equiv 0, \ee since no
deformation is present in the UV. So only the second term of the
solution enters the closed string overlap. We have thus to compute
\be
Tr[V_c\,\psi_u]&=&-Tr[V_c\,\frac1{K+\phi_u}u\del_u\phi_uBc\del c]\0\\
&=&
-\int_0^\infty dT\left\langle V_c(i\infty)Bc\del c(0)\, u\del_u\phi_u(0)e^{-\int_0^T ds \phi_u(s)}\right\rangle_{C_T}\0,
\ee
rescaling $C_T\to C_1$ (the closed string is scale invariant since it is on shell) and changing variable $y=Tu$ we obtain an explicit $u$--independent quantity (which explicitly shows that $u$ is a gauge redundancy)
\be
Tr[V_c\,\psi_u]&=&-\int_0^\infty dT\,u\left\langle V_c(i\infty)Bc\del c(0)\, \del_{Tu}\phi_{Tu}(0)e^{-\int_0^1 ds \phi_{Tu}(s)}\right\rangle_{C_1}\0\\
&=&-\int_0^\infty dy\,\left\langle V_c(i\infty)Bc\del c(0)\,
\del_{y}\phi_{y}(0)e^{-\int_0^1 ds
\phi_{y}(s)}\right\rangle_{C_1}.\0 \ee Now we have to get rid of the
$B$ line. Remembering that
$$V_c(i\infty)=c(i\infty)c(-i\infty)V_{matter}(i\infty,-i\infty),$$
and since $B$ vanishes (fast enough) at the midpoint, we just
have\footnote{See \cite{Erler:2007xt} for a derivation of this,
where the role of the closed string insertion is played by the
picture changing operator, whose $c$ content is the same as $V_c$.}
\be \left\langle
V_c(i\infty)Bc(t_1)c(t_2)\right\rangle_{C_T}=\frac{t_1}{T}\left\langle
V_c(i\infty)c(t_2)\right\rangle_{C_T}-\frac{t_2}{T}\left\langle
V_c(i\infty)c(t_1)\right\rangle_{C_T}. \ee We thus obtain \be
Tr[V_c\,\psi_u]&=&+\int_0^\infty dy\,1\times\left\langle V_c(i\infty)c(0)\, \del_{y}\phi_{y}(0)e^{-\int_0^1 ds \phi_{y}(s)}\right\rangle_{C_1}\0\\
&=&\int_0^\infty dy\,\left\langle V_c(i\infty)c(0)\,(- \del_{y})e^{-\int_0^1 ds \phi_{y}(s)}\right\rangle_{C_1}\0\\
&=&\int_0^\infty dy\,(- \del_{y})\left\langle V_c(i\infty)c(0)\,e^{-\int_0^1 ds \phi_{y}(s)}\right\rangle_{C_1}\0\\
&=&\left(\lim_{y\to 0^+}-\lim_{y\to \infty}\right)\left\langle V_c(i\infty)c(0)\,e^{-\int_0^1 ds \phi_{y}(s)}\right\rangle_{C_1}\0\\
&=&\ll V_c\gg^{UV}-\ll V_c\gg^{IR},
\ee
as wanted. Notice the importance of (\ref{cnd3}) in order to assure that $\ll V_c\gg^{IR}\neq0$, so that we are not computing a tachyon vacuum observable.

\section{Energy, a general expression for the cubic term}

Assuming the validity of the equation of motion against the solution itself, the energy of a time independent classical solution is given by
\be
E(\psi)=-S[\psi]=\frac16Tr[\psi Q\psi]=-\frac16Tr[\psi^3]\label{cubic-nrg}
\ee
Since two Schwinger parameters are needed to represent $Q\psi$ as a superposition of deformed wedges, we do not get any simplification in computing $Tr[\psi_u Q\psi_u]$ rather then $Tr[\psi_u^3]$. This is because, contrary to the universal tachyon vacuum case, the deformed wedge states are not BRST closed.
Under the assumption $(c\phi_u)^2=0$, the identity piece in the solution drops from the evaluation of the energy, and we are left with
\be
Tr[\psi_u^3]&=&-Tr\left[\frac1{K+\phi_u}u\del_u\phi_u Bc\del c\,
 \frac1{K+\phi_u}u\del_u\phi_u Bc\del c\,\frac1{K+\phi_u}u\del_u\phi_u Bc\del c \right]\0\\
&=&-\int_0^\infty dt_1dt_2dt_3\left\langle e^{-\int_0^Tds\phi_u(s)}Bc\del c
u\del_u\phi_u(T(x+y))\,\del c u\del_u\phi_u(T x)\,
\del c u\del_u\phi_u(0)\right\rangle_{C_T}\0\\
&=&-\int_0^\infty dT\,u^3T^2\int_0^1 dx\int_0^{1-x} dy\0\\
&&\quad\quad\quad \cdot\left\langle
e^{-\int_0^Tds\phi_u(s)}Bc\del c \del_u\phi_u(T(x+y))\,\del c
\del_u\phi_u(T x) \,\del c \del_u\phi_u(0)\right\rangle_{C_T},\0
\ee
where we have defined (and accordingly changed variables in the
integration) \be
T&\equiv& t_1+t_2+t_3\\
x&\equiv& \frac{t_1}{T}\\
y&\equiv& \frac{t_2}{T}. \ee
Now we rescale $C_T\to C_1$ and get
\be
Tr[\psi_u^3]&=&-\int_0^\infty dTu^3T^2\!\int_0^1\! dx\int_0^{1-x}
\!dy\0\\
&&\quad\quad\cdot
\left\langle e^{-\int_0^1ds\phi_{Tu}(s)}Bc\del c
\del_{Tu}\phi_{Tu}(x+y) \,\del c \del_{Tu}\phi_{Tu}( x)\,\del c
\del_{Tu}\phi_{Tu}(0)\right\rangle_{C_1}.\0
\ee
As a last step we
change variable $$q=uT$$ to get an explicitly $u$--independent
expression
\be
Tr[\psi_u^3]&=&-\int_0^\infty dq\,q^2\int_0^1 dx\int_0^{1-x} dy \,F(q,x,y)\0\\
F(q,x,y)&=&\left\langle e^{-\int_0^1ds\phi_{q}(s)}Bc\del c
\del_{q}\phi_{q}(x+y)
\,\del c \del_{q}\phi_{q}(x)\,\del c \del_{q}\phi_{q}(0)\right\rangle_{C_1}\0\\
&=&\left\langle Bc\del c(x+y) \,\del c(x) \,\del c(0)\right\rangle^{(gh)}_{C_1}\0\\
&\times&\left\langle e^{-\int_0^1ds\phi_{q}(s)}
\del_{q}\phi_{q}(x+y)\, \del_{q}\phi_{q}(x)
\,\del_{q}\phi_{q}(0)\right\rangle^{(matt)}_{C_1}.\label{nrgcorr}
\ee

In the case of a universal tachyon vacuum solution $\phi_q=q$, as it
is easy to check, we get back Erler--Schnabl result. But in general,
although the structure seems promising, we do not know how to
compute the 3-point function in the deformed matter correlator in
full generality: for the closed string overlap the matter correlator
was just a one--point function in the deformed theory, which could
be easily obtained by differentiating the partition function. Here
there are three  matter+ghost insertions, which upon integration
should still give a result related to an appropriate differential
operator acting on the partition function. In fact, at the end of
the day the correct result should be given by \be -\frac16
Tr[\psi_u^3]=-\frac{1}{2\pi^2}(g(u=0)-g(u=\infty))=-E^{(UV)}+E^{(IR)},
\ee where $g(u)$ is the open string partition function on the
canonical cylinder \be g(u)=\left\langle
e^{-\int_0^1ds\phi_{u}(s)}\right\rangle_{C_1}. \ee This is because,
assuming $\frac {g(u=\infty)}{g(u=0)}=\frac {E^{(IR)}}{E^{(UV)}}$,
\cite{tension}, and knowing that for  tachyon vacuum universal
solutions we have
\be
E(\psi_{TV})=-E^{(D25)}=-\frac1{2\pi^2}\bra0\ket0^{matt}=-\frac1{2\pi^2}\,\langle
1\rangle=-\frac1{2\pi^2}\,(g(0))^{26}=-\frac1{2\pi^2}\prod_{i=0}^{25}\left(\frac{V_i}{2\pi}\right),\0
\ee
this is the only way the correct ratio of tensions can be
obtained.

\section{Examples}

In the previous section we constructed a solution based on the seed $\phi_u$ subject to the conditions
\be
(c\phi_u)^2&=&0\\
f_t\circ\phi_u(z)&=&\frac1t\phi_{tu}\left(\frac zt\right)\\
\phi_u-\delta\phi_u&=&u\del_u\phi_u.
\ee
In addition, in order to flow to a well defined IR limit (different from the tachyon vacuum), the
partition function
\be
g(u)\equiv\left\langle e^{-\int_0^1ds\phi_u(s)}\right\rangle_{C_1},
\ee
should obey
\be
\lim_{u\to\infty}g(u)=Z_{IR}=\verb"finite".
\ee
This last constraint, which is also the most important one, is the only one that we cannot satisfy without knowing in advance the exact RG flow on the
worldsheet. We now show how some well--known worldsheet RG flows naturally fall inside this scheme.

\subsection{An explicit example: lumps in non-compact dimensions}

A remarkable example (which only lives in the uncompactified limit) is the famous Witten \cite{Witten92} boundary deformation.
This deformation was considered by many authors in the context of Boundary String Field Theory, \cite{Kutasov}, and, more recently, by Ellwood, \cite{Ellwood:2009zf} in a pioneering
proposal for Lump Solutions. The deformation (defined in the $\frac2\pi\arctan$--sliver frame is given by)
\be
\phi_u(z)&\equiv&uf_u\circ(:X^2:(uz)+A)=u[:X^2:(z)+2(\log u+A)],\\
f_u(z)&=&\frac{z}{u}.
 \ee
The key property of such insertion is that, when exp-integrated on
the boundary it interpolates between Neumann ($u=0$) and Dirichlet
($u=\infty$), \cite{Witten92}.

The $u\log u$ piece is there to compensate for the non--primary
conformal transformation of $:X^2:$ \be
f\circ:X^2(z):=:X^2(f(z)):-\log|f'(z)|^2. \ee This implies that
under conformal transformations, the parameter $u$ gets changed as follows
\be
f\circ\phi_u(z)=\frac1{|\partial_{z'}f^{-1}(z')|}\phi_{u|\partial_{z'}f^{-1}(z')|}(z').
\ee In particular, under scaling $f_t(z)=z/t$, we have \be
f_t\circ\phi_u(z)=\frac1t\phi_{tu}\left(\frac zt\right) \ee and
consequently \be f_t\circ\int_a^bdy\,\phi_u(y)=\int_{\frac
at}^{\frac bt}d\tilde y\,\phi_{tu}\left(\tilde y\right). \ee The
pure number $A$, which depends on the normal ordering conventions
(we choose the standard UHP convention), must be
determined in order to have a finite partition function in the infrared, see later.\\
Now we just need to derive $\delta\phi$ by making the BRST variation
\be [Q,\phi_u(z)]= u[Q,:X^2(z):]=u\left(-2\del
c(z)+c\del:X^2(z):\right). \ee
The second piece in the BRST
variation does not enter the solution and the first piece (the one
proportional to $\del c$) defines $\delta\phi$ to be \be
\delta\phi_u(z)=-2u. \ee
Notice that this is just a number.\\
In total, one can check  that (\ref{cnd2}) is satisfied
\be
\phi_u-\delta\phi_u=u\del_u\phi_u.
\ee

$A$ is a $u$--independent constant which must be determined in order
for the partition function to be finite in the $u\to\infty$ limit
(that is the IR fixed point of the world--sheet boundary deformation). In
Appendix D we review this derivation and explicitly find \be
A=\gamma-1+\log4\pi.\label{A} \ee We remark that A is not just an
anonymous integration constant, but represents 'the right amount' of
zero momentum tachyon that is necessary to insert in order to get a
finite partition function in the IR.

Integrating $\phi_u$ on a cylinder $C_T$ we get the partition function which is given by, see Appendix D,
\be
\left\langle e^{-\int_0^Tds\phi_u(s)}\right\rangle_{C_T}&=&\left\langle e^{-\int_0^1 ds\phi_{Tu}(s)}\right\rangle_{C_1}=g(Tu)\\
g(u)&=&\frac1{\sqrt{2\pi}} \sqrt{2u}\,\Gamma(2u)\left(\frac e{2u}\right)^{2u}.
\ee
In the UV limit $u\to 0$ we have
\be
\lim_{u\to0} g(u)=\lim_{u\to0} \frac1{2\sqrt{\pi u}}=\delta(0)=\int_{-\infty}^{\infty}\frac {dx}{2\pi}=\bra 0\ket0=\frac{V_1}{2\pi},
\ee
while for $u\to\infty$
\be
\lim_{u\to\infty} g(u)=\lim_{u\to\infty} \,e^{-2u(\gamma-1+\log4\pi-A)}\left(1+O\left(\frac 1u\right)\right){\Big|}_{A=\gamma-1+\log4\pi}=1.
\ee
This indeed reproduces the ratio of tensions (using the fact that the total mass, up to some universal constant, is proportional to the partition function, \cite{tension})
\be
\frac{\tau^{(D24)}}{\tau^{(D25)}}=\frac{g(\infty)}{\frac{g(0)}{V_1}}=\frac1{\frac1{2\pi}}=2\pi.
\ee

Finally we would like to remark that it is not hard to construct
D(25-p)-branes, by introducing additional coordinate fields and couplings, starting
from $\sum_{i=1}^p u_i :\!X^2_i\!:(z)$,  instead of the simple $u :\!X^2\!:(z)$ as above. Indeed we know from \cite{Witten92,Kutasov} that the renormalization group flow does not mix up such different couplings, which continue to evolve linearly. As a consequence the construction of D23, D22,...--brane solutions is a straightforward generalization of the one for the D24-brane, see \cite{BGT2} for the explicit construction.

\subsubsection{Multiple lower dimensional branes?}

We saw that having a quadratic minimum in the worldsheet boundary integration places a lower dimensional D-brane at that minimum. A simple generalization
consists in taking a $\phi_u$ which has several quadratic minima\footnote{It appears natural to require the minima to be at the same height although, without
 explicitly solving the RG flow near the IR, it is not possible to be more concrete on this point.}. Then, according to the heuristic property that in the IR the X field is constrained to stay at the minimum on the boundary potential, we expect to have several displaced $D24$-branes. Here we just want  to point out that, as far as the general structure is concerned, our solution can easily handle such a non--abelian configuration. Concentrating on a case with two distinct quadratic minima we can consider in full generality the relevant field
\be
\chi^{(a)}(z)=:(X-a)^2(X+a)^2:=:X^4:-2a^2:X^2:+a^4,\quad a>0
\ee
The corresponding seed can be taken to be ($f_t(z)=\frac zt$)
\be
\phi_u^{(a)}&=&uf_u\circ \chi^{(a)}(uz)\0\\
&=&X^4(z)+(12\log u-2a^2)X^2(z)+12\log^2u-4a^2\log u+a^4,
\ee
where use have been made of (normal ordering is understood)
\be
f\circ X^4(z)&=&(-i\del_p)^4 f\circ e^{ipX(z)}{\Big|}_{p=0}=\del^4_p\left(|f'(z)|^{p^2}e^{ipX(f(z))}\right)\0\\
&=&X^4(f(z))-12\log|f'(z)|X^2(f(z))+12\log^2|f'(z)|. \ee Since
$f_t\circ f_u=f_{tu}$, the seed defined in this way will
automatically satisfy \be
f_t\circ\phi_u(z)=\frac1t\phi_{tu}\left(\frac zt\right). \ee using
\be \delta X^n=(-i\del_p)^n\delta
e^{ipX}{\Big|}_{p=0}=(-i\del_p)^n\,p^2 e^{ipX}{\Big|}_{p=0}, \ee one
can also verify that \be
\phi_u^{(a)}-\delta\phi_u^{(a)}=u\del_u\phi_u^{(a)}. \ee
It is is
easy to see that contact term divergences are still logarithmic,
which is more than enough to make \be (c\phi_u^{(a)})^2&=&0 \ee and
\be Tr[e^{-T(K+\phi_u^{(a)})}(...)]\to\langle
e^{-\int_0^Tds\phi_u^{(a)}(s))}(...)\rangle_{C_{T+(...)}} \ee
without need of regularization/renormalization. This can be of
course generalized to coincident branes (just take $a=0$), and the
number of branes can be straightforwardly increased starting with
$:X^{2n}:$ and `covariantizing' it by defining $$\phi_u(z)=uf_u\circ
:X^{2n}:(uz).$$ Unfortunately, it is only for the $X^2$ deformation
that we can completely solve the worldsheet theory (since in that
case it is just a free theory for the $X$ field, with a boundary
`mass' term). In particular, without knowing the partition function
in the infrared, we cannot tune the seed $\phi_u$ in order to make
$\frac1{K+\phi}$ singular while $Tr[\frac1{K+\phi}(\phi-\delta\phi)]$
finite, so this remains as a proposal for further investigations.
\subsection{Other examples: primary seeds}
The simplest example of a seed satisfying the conditions
(\ref{cnd1},\ref{cnd2}) is actually given by taking a primary
relevant operator $\chi^{(h)}(z)$ of weight $h<\frac12$ (in order to
avoid contact term divergences in the solution) and to define \be
\phi_u(z)=u (f_u\circ
\chi^{(h)}(uz)+A(h))=u(u^{-h}\chi^{(h)}(z)+A(h)). \ee Condition
(\ref{cnd1}) is satisfied because \be
f_t\circ\chi^{(h)}(z)=t^{-h}\chi^{(h)}(z/t), \ee Condition
(\ref{cnd2}) is satisfied because \be
(1-\delta)\chi^{(h)}(z)=(1-h)\chi^{(h)}(z). \ee The pure number
$A(h)$ (which will in general depend on the weight $h$) should be
determined in order for the IR partition function to be finite. This
requires to be able to exactly solve the worldsheet theory in the
presence of the boundary interaction given by $\int_{\del M} ds
\chi^{(h)} (s)$. This is not known in general except for a few
integrable interactions.

 The prototype example of this is the
celebrated $\cos(X/R)$ deformation (with $R>\sqrt{2}$, because of
the technical assumption $(c\phi)^2=0$), which is known to be
exactly solvable, \cite{Fendley:1994rh, Fendley}. In this case the
solution will describe the flow between a D25 brane wrapped on a
circle of radius R and a D24--brane placed at the minimum of the
boundary potential.
 Let us see in this case how to tune the seed in
order to get a finite partition function in the infrared, thus
obtaining a solution describing the condensation of a brane wrapping
a circle to a lower dimensional brane living at a specific point of
the circle.

 In \cite{Fendley}, first using TBA techniques
\cite{Fendley:1994rh}, and then checking in conformal perturbation theory,
the following IR limit was derived (which we write down using
standard `String (Field) Theory' convention)
\be \frac{\left\langle
e^{\lambda\int_0^T\,\cos\frac{X(s)}{R}}
\right\rangle_{C_T}}{\left\langle1  \right\rangle_{C_T}}\sim
\frac1R\exp\left[T\left(\frac
\lambda{2k'}\right)^{\frac{R^2}{R^2-1}}\frac1{2\cos\frac{\pi}{2(R^2-1)}}\right],\quad\quad
\lambda\to\infty \ee where $k'$ is some positive constant (relating
the coupling $\lambda$ to the boundary temperature,
\cite{Fendley}). Notice that the above expression
is well defined for $R^2>2$, which is the usual condition which makes the ordered exponential finite without the need of regularization/renormalization.\\
That's all we need.

In our setting we define the un--tuned seed \be \hat\phi_u\equiv
uf_u\circ\left(-\cos\frac XR\right)=-u^{\frac{R^2-1}{R^2}}\cos\frac
XR, \ee and we are interested in the large $u$ limit of \be
\frac{\left\langle e^{-\int_0^T\,\hat\phi_u(s)}
\right\rangle_{C_T}}{\left\langle1  \right\rangle_{C_T}}. \ee By
comparison we have \be \lambda=u^{\frac{R^2-1}{R^2}}, \ee which
readily gives \be \frac{\left\langle e^{-\int_0^T\,\hat\phi_u(s)}
\right\rangle_{C_T}}{\left\langle1
\right\rangle_{C_T}}\sim\frac1R\exp\left[\eta(R)Tu\right],\quad\quad
u\to\infty \ee where we have defined \be
\eta(R)=\frac1{(2k')^{2\,\frac{R^2}{R^2-1}}\,\cos\frac{\pi}{2(R^2-1)}}.
\ee Notice that with our choice of seed the RG--flow is exactly
linear ($u$ increases linearly under scaling), which is a quite non
trivial consistency check of the condition (\ref{cnd1}).

Now we tune the seed by adding the appropriate contribution proportional to
the identity operator
\be
\phi_u=\hat\phi_u+u A(R),
\ee
and we uniquely determine $A(R)$ in order for the partition function to be finite in the IR
\be
A(R)=\eta(R).
\ee
We then have
\be
\lim_{u\to\infty}\frac{\left\langle e^{-\int_0^T\,\phi_u(s)}  \right\rangle_{C_T}}{\left\langle1  \right\rangle_{C_T}}=\frac1R=\frac{Z^{IR}}{Z^{UV}}.
\ee
This, of course, reproduces the ratio of tensions
\be
\frac{\tau^{D24}}{\tau^{D25}}=Vol\frac{Z^{IR}}{Z^{UV}}=2\pi R\frac1R=2\pi.
\ee

We would like to point out that the  necessity of tuning the seed solves a
puzzle raised in \cite{Harvey} about the fact that both $\cos$ and $-\cos$
give the same (but translated) D24-branes. This, in turn, looked inconsistent
 with the non--symmetric shape of the OSFT tachyon potential under $T\to-T$.
 But from the above
analysis it is evident that, after adding the $A(R)$--correction (which, being
 a number, is usually ignored in the world--sheet analysis as it corresponds
 to an additive constant in the world--sheet action), the $T\to-T$
 asymmetry is restored and $-\phi_u$ will definitely give a divergent
  partition function in the infrared, telling us that the solution is falling
  in the unbounded side of the tachyon potential.

\subsubsection{Interpolating from infinite to finite volume}

It is useful to understand the infinite transverse volume  as a limit of a circle with increasing radius $R$. Since the $:X^2:$ deformation only lives in the
strict $R=\infty$ limit, it cannot be used on the circle. In order to define a deformation of it which can live on circles, we observe that
\be
:X^2:=-\del_p^2 :e^{ipX}:{\Big |}_{p=0}.
\ee
Using
\be
\del_p^2 f(p){\Big |}_{p=0}=\lim_{\epsilon\to0}\frac{f(\epsilon)-2f(0)+f(-\epsilon)}{\epsilon^2},
\ee
and the fact that on a circle of radius $R$ we have $\epsilon=\frac1R$, because of momentum quantization we get that $X^2$ is deformed to
\be
:X^2:\quad\rightarrow\quad 2R^2\left(1-:\cos\frac XR:\right)=:X^2:+O\left(\frac1R\right).
\ee
The corresponding seed will thus be given by
\be
\phi_u^{(R)}&\equiv& uf_u\circ\left(2R^2\left(1-:\cos\frac XR:\right)+A(R)\right)\\
&=&-2uR^2\left(u^{-\frac1{R^2}}:\cos\frac XR:-1-\frac {A(R)}{R^2}\right).\0
\ee
Again, the quantity $A(R)$ should be derived by studying the partition function in the $u\to\infty$ limit. Using the result of the previous paragraph we get
\be
A(R)=2R^2-\frac1{2\cos\frac{\pi}{2(R^2-1)}}\left(\frac{R^2}{k'}\right)^{\frac {R^2}{R^2-1}},\label{ARR}
\ee
and the ratio of tensions is obviously reproduced. The precise matching with the $A$ determined in the $R=\infty$ limit, (\ref{A}), would require some  information about the unknown parameter $k'$ and its possible (mild) dependence on R (in order to cancel the subleading logarithmic divergence of (\ref{ARR}) in the $R\to\infty$ limit).

\section{Conclusion and discussion}

The content of our paper can be summarized as follows: {\it given a worldsheet boundary RG flow, generated by the boundary integration of a relevant matter field
(whose contact term divergences are mild enough to make $e^{-\int_a^bds\,\phi(s)}$ finite without the need of renormalization), we are able to associate to it a simple solution of  OSFT  representing the new BCFT which is met at the
end of the RG flow.}\\
The solution is defined by the two crucial quantities
\be
&&\frac1{K+\phi}\0\\
&&\frac B{K+\phi}(\phi-\delta\phi).\0 \ee If the first quantity is
regular, then the cohomology of the solution can be trivialized by
the well-defined homotopy field \be A_\phi=\frac B{K+\phi}, \ee and the
solution will  describe the tachyon vacuum. Thus a singular
$\frac1{K+\phi}$ is needed in order for the solution to describe an
IR fixed point where the appropriate open string dynamics is left.
This is controlled by the second quantity which, since it enters
explicitly in the solution, should be finite in correlators. These two conditions are needed to describe a non trivial IR fixed point.\\

In principle $any$ $\phi$ such that $\frac1{K+\phi}$ is singular while
 $\frac B{K+\phi}(\phi-\delta\phi)$ is finite, can be used to identify a
 new BCFT, the IR fixed point. However, it is in general not easy to `tune'
  $\phi$ so that the two conditions are simultaneously met.

One of the main result of the paper is the recognition that, for a
simple and `natural' class of seeds (\ref{cnd1},\ref{cnd2}), these
two conditions boil down to the requirement of a finite ( and non
vanishing)  partition function in the infrared. We indeed find that
the trace of $\frac1{K+\phi}(\phi-\delta\phi)$ is nothing but the
shift in the partition function from the UV to the IR \be
Tr[\frac1{K+\phi}(\phi-\delta\phi)]=Z_\phi^{(UV)}-Z_\phi^{(IR)}. \ee
Thanks to the same mechanism the Ellwood invariants of the solution
will correctly compute the corresponding shift in the closed string
one--point function.

 It would be nice to explicitly compute the
boundary state of the solution \cite{Kiermaier:2008qu} to verify whether
\be \ket
{B_*(\psi_u)}=?=\lim_{u\to\infty}e^{-\int_0^{2\pi}\frac{d\theta}{2\pi}
\phi_{u}(\theta)}\ket B. \ee This is expected since the solution is
still wedge--based and an explicit check would be instructive, especially in view of the results presented in \cite{EM}.

For the energy things are not so simple: already for the simplest
relevant deformation given by the tachyon vacuum the energy equals
the shift in the partition function thanks to a kind of `miracle'.
In fact, after integrating on the first two Schwinger parameters, the energy of the Erler--Schnabl tachyon vacuum turns out to be
 \be
E(\psi_{TV})&=&-\frac16Tr[\psi_{TV}^3]=-\frac1{2\pi^2}\int_0^\infty dt \frac12 t^2e^{-t}\0\\
&=&\frac1{2\pi^2}\int_0^\infty dt \del_t F(t)=\frac1{2\pi^2}(F(\infty)-F(0)),
\ee
where
\be
F(t)&=&\frac12(t^2+2t+2)g(t)\0\\
g(t)&=&e^{-t}=\langle e^{-t\int_0^{2\pi} \frac{d\theta}{2\pi}}\rangle_{Disk}=\texttt{Partition Function}.
\ee
The energy then equals the shift in the partition function because
\be
F(0)&=&g(0)=<1>=\frac {V^D}{(2\pi)^D}\0\\
F(\infty)&=&g(\infty)=0.
\ee
In other words the on-shell action is
the integral of a total derivative of a function which equals the
partition function only `on shell' (that is at the conformal fixed
point). While this is basically what happens in Boundary String
Field Theory, we have not set out in the present paper to compute
the `off shell' partition function $F(t)$ for a general relevant
deformation. We think however this will be possible by carefully
analyzing the involved matter+ghost correlator, (\ref{nrgcorr}).

 Notice that, just
because we correctly described the coupling of on--shell closed
strings to our solution (and because such coupling is always
proportional to the tension), the correct ratio of tensions has been
already reproduced. But, needless to say, the consistency check that
the ratio
is the same from the computation of the on-shell action is missing, see \cite{EM} and \cite{BGT}.\\
What is also lacking is a careful analysis of the cohomology around
the solution, see \cite{EM}.\\

 Notwithstanding the many important aspects that still need to be
 refined and understood in detail, we believe that we provided the
 first explicit proposal for an analytic solution of OSFT representing a relevant
 boundary deformation.

We end with a couple of comments. Since very few boundary RG flows
are explicitly computable, what we really would like to do is to
{\it use} OSFT to {\it define} the new BCFT which is the result of
the (unknown) RG--flow, as it was numerically done in \cite{lumps}.
Unfortunately we do not have very concrete things to say about this.
The problem is that, no matter  what the relevant operator is that
we start with, it is extremely easy to just flow to the tachyon
vacuum, or to get a singular solution. The reason for this is
essentially the fact that sliver--like projectors (which are the ones
that, upon rescaling, give the partition function in the IR) are
defined as infinite products of `wedge states' and, as such, are
extremely sensitive to the normalization in front of such wedge
states. In our language this means that the relevant operator we use
should be supplied with a part proportional to the identity
operator, that should be fixed (uniquely) by the requirement of a
finite partition function in the infrared. This in turns require to
be able to calculate the behavior of the RG flow near the IR. It
should be  stressed that, ultimately, this problem is linked to the
definition of well--defined  sliver projectors, rather than the
solutions themselves.\\
 It would be certainly desirable to analyze the solution we presented in level expansion, in a convenient conformal frame. However, a genuine $L_0$ (such as the one of
\cite{lumps})or $\mathcal L_0$ level expansion  would be well defined only in a
compactified setting in order to have a discrete spectrum for the
momentum, which would oblige us to discard the best understood
relevant deformation based on $X^2$. \\
\no One can also wonder if our solution can be  obtained by fixing a suitable gauge, and then iteratively solving the equation
of motion starting with a relevant field and, correcting it order by
order, to end up with a finite solution, as it was done in \cite{expsft}, for the tachyon vacuum. It appears natural that the seed itself will enter the gauge fixing
condition, which will be a sort of generalization of the `dressed'
$B$--gauge fixing conditions of \cite{ErlerSchnabl}.

We think these are  important aspects to be
 addressed in the future.

\acknowledgments

C.M. thanks Ted Erler, Leonardo Rastelli and Martin Schnabl for interesting discussions.
We would like to thank the organizers of the workshop �APCTP Focus Program on Current Trends in String Field Theory� at APCTP, Pohang, South Korea, where this work was started, for the hospitality and for the stimulating environment.
\noindent C.M. would like to thank SISSA and the Simons Workshop in Mathematics and Physics 2010 for the kind hospitality
during part of this research. The work of D.D.T. was supported by
the Korean Research Foundation Grant funded by the Korean Government
with grant number KRF 2009-0077423. The research of L.B. was
supported in part by the Project of Knowledge Innovation Program
(PKIP) of Chinese Academy of Sciences, Grant No. KJCX2.YW.W10.

\section*{Appendix}
\appendix
\section{Building the solution from the Tachyon Vacuum}
Here we show that our solution can be alternatively obtained following the construction proposed by Ellwood, \cite{Ellwood:2009zf}. We start with a `seed' string field $\Gamma$ which is not a solution and use it to construct
\be
\Omega_L&=&-A\Gamma\\
\Omega_R&=&-\Gamma A,
\ee
where $A$ is the homotopy field at the tachyon vacuum
\be
{\cal Q}A=1.
\ee
We can then use $\Omega_L$ and $\Omega_R$ to build two L/R solutions at the tachyon vacuum
\be
\tilde\psi_L&=&\frac1{1-\Omega_L}{\cal Q} (1-\Omega_L)=U_L^{-1}{\cal Q} U_L\\
\tilde\psi_R&=&(1-\Omega_R){\cal Q} \frac1{1-\Omega_R}= U_R{\cal Q}
U_R^{-1}. \ee If $\Gamma$ were a solution, then we would have
$\tilde\psi_L=\tilde\psi_R=\Gamma$. However $\tilde\psi_{L,R}$ are
different if $\Gamma$ is not a solution. It is useful to trace them
back to the perturbative vacuum, where they will also be expressed
in pure gauge form. \be
\psi_L&=&(U_L^{-1}U_0)Q(U_0^{-1}U_L)=V_L Q V_L^{-1}\\
\psi_R&=&(U_R U_0)Q(U_0^{-1}U_R^{-1})=V_R^{-1} Q V_{R}.
\ee
The tachyon vacuum solution is given by
\be
\psi_0=U_0QU_0^{-1}
\ee
where
\be
U_0=1-hFBcFh^{-1}
\ee
$h=h(K)$ is a gauge freedom whose only effect is to change the security strips of the tachyon vacuum solution which is\footnote{To be precise, the gauge
freedom given by $h$, changes the tachyon vacuum so, in general, $\tilde\psi_{L,R}$ will be solutions around two different, gauge equivalent forms, of the tachyon
vacuum. This subtlety is anyhow unimportant when the solutions are brought back to the perturbative vacuum $\psi=0$, as we do.}
\be
\psi_0=hF\,c\,\frac{KB}{1-F^2}\,c\,Fh^{-1}
\ee
choosing $h=F^{-1}$ we can put the  strip on the right, while for
$h=F$ it will be on the left. The homotopy field is independent of $h$ and is
given by
\be
A=B\,f,
\ee
where we have defined for convenience
\be
f=f(K)=\frac{1-F^2(K)}{K}.
\ee
This is a very general scheme which is not guaranteed to work in general.
Let us now test it for the $D25$ brane.

\subsection{Seeding the D25-brane}

We begin by showing how to get the D25 brane solution, starting from an {\it identity--based} seed which is simply
\be
\Gamma=-c
\ee
which gives
\be
\Omega_L&=&fBc\\
\Omega_R&=&cBf.
\ee
Following the scheme of the previous section we write directly the solutions at the $perturbative$ vacuum.
Concentrating on the left solution
\be
\psi_L=(U_L^{-1}U_0)Q(U_0^{-1}U_L)=V_L Q V_L^{-1},
\ee
 we get
\be
V_L=1+\frac{f}{1-f}Bc-\frac1{1-f}\,hF\,Bc\,Fh^{-1}.
\ee
Now we use the gauge freedom of $h$ to write the transformation in canonical form
\be
h&=&F\\
V_L&=&1+\frac{f-F^2}{1-f}Bc\\
V_L^{-1}&=&1-\frac{f-F^2}{1-F^2}Bc.
\ee

In general, we have a non trivial solution (at the $perturbative$ vacuum) when $V_L^{-1}$ is singular. In this case
it is easy to see that for any ``good'' $F(K)$ (that is $F(K)=1+\alpha K +O(K^2)$ around $K=0$) both $V_L$ and $V_L^{-1}$ are well defined. So this solution
is genuinely gauge equivalent to the perturbative vacuum. This becomes even more transparent if we write the tachyon vacuum solution in the gauge
$$F^2=f, $$
which uniquely determines
\be
F^2(K)=\frac 1{1+K},
\ee
that is the Erler--Schnabl tachyon vacuum  solution. In this gauge we have explicitly
\be
V_L&=&V_L^{-1}=1\\
\psi_L&=&0.
\ee

To find the  `right' solution we proceed in the same way but now we put the external strip of the TV solution to the right.
That is we choose $h=F^{-1}$. With straightforward algebra we can write
\be
\psi_R&=&V_R^{-1}QV_R\\
V_R&=&\frac1{1-F^2}\left(1+cB\frac{f-F^2}{1-f}\right)\rightarrow\left(1+cB\frac{f-F^2}{1-f}\right)\\
V_R^{-1}&=&\left(1-cB\frac{f-F^2}{1-F^2}\right)(1-F^2)\rightarrow\left(1-cB\frac{f-F^2}{1-F^2}\right).
\ee
The $K$--dependent factors $(1-F^2)^{\pm1}$, although singular, are $reducible$ gauge transformations (they do not act) and can thus be dropped.
Notice that the right quantities (modulo the reducible gauge transformations) are obtained by reading the left ones form right to left and viceversa.
Again it is obvious that both $V_R$ and $V_R^{-1}$ are well defined, so this is the perturbative vacuum.

\subsection{Seeding a lower dimensional brane}

To construct a lower dimensional brane we choose the identity--based seed
\be
\Gamma=-c\tphi,
\ee
where $\tphi$ is some relevant matter operator.
This gives
\be
\Omega_L&=&fBc\tphi\\
\Omega_R&=&\tphi cBf,
\ee
In order to get the simplest possible solution we choose Erler-Schnabl Tachyon Vacuum since the beginning.
\be
F^2=f=\frac1{1+K}.
\ee
Then the gauge transformations at the perturbative vacuum are given by
\be
V_L&=&1-\frac{1}{1+K-\tphi}(1-\tphi)Bc\\
V_L^{-1}&=&1+\frac{1}{K}(1-\tphi)Bc,
\ee
which, after the redefinition
\be
\tphi=1-\phi,
\ee
gives the gauge transformation (\ref{gaugetransf}) of section 3.2.
\no For completeness we also show  the solutions $\psi_L$ and $\psi_R$ in arbitrary $F^2\neq f$ gauge,
 \be
 \psi_L=&\frac{1}{1-f\tphi}(F^2-f\tphi)\Big(cKBc+c\frac{K}{1-F^2}(F^2-f\tphi)Bc\Big)\0\\
 &+\frac{1}{1-f\tphi}\Big(f[\tphi,K]c-f\partial
 c \delta\tphi Bc\Big)
 \ee
and \be
 \psi_R=&\Big(cBKc+cB(F^2-\tphi f)\frac{K}{1-F^2}c\Big)(F^2-\tphi f)\frac{1}{1-\tphi f}\0\\
 &+\Big(c[K,\tphi]f+cB\partial
 c \delta\tphi f\Big)\frac{1}{1-\tphi f}
 \ee
 Still $\psi_R$ is obtained by reading $\psi_L$ from left to right, but when $F^2\neq f$ these solutions do not look simple at all. This is another way of
 appreciating the `simplicity'  of our solution.

\section{Explicit real form for the energy}

The energy of a static solution $\psi$ is measured by (is proportional to) the expression $\langle \psi Q \psi\rangle$, or alternatively by $-\langle \psi \psi \psi\rangle $. We find it more convenient to use the latter.
\be
E= -\frac 16 \langle \psi \psi \psi\rangle \label{solenergy}
\ee
 To  simplify the notation let us set
\be
X=\frac 1{K+\phi},\quad\quad H=\phi-\delta \phi\label{XH}
\ee
so that
\be
\psi_L = c\phi- XHBcKc, \quad\quad \psi_R=  c\phi-HcKcB X
\ee
Of course $B$ commutes with $\phi,X$ and $H$. We have in addition
\be
[B,cKc]=[K,c],\quad\quad \{B,[K,c]\}=0,\quad\quad BcKcB=[K,c]B\label{LR0}
\ee
and
\be
\{c,[K,c]\}=0 ,\quad\quad [c,\phi]=0,\quad\quad [c,H]=0,\quad\quad [cKc,H]=0\label{LR0'}
\ee
It is easy to prove that
\be
&&\langle \psi_L \psi_L \psi_L \rangle =- \langle XHBcKc\,\,XHBcKc\,\, XHBcKc\rangle \0\\
&&= - \langle HBcKc\,\, XHBcKc\,\, XHBcKcX\rangle=  - \langle HcKcXB\,\, HcKcXB\,\, HcKcXB\rangle\0\\
&&= -\langle HcKcBX\,\, HcKcBX\,\, HcKcBX\rangle= \langle \psi_R \psi_R \psi_R \rangle \label{L=R}
\ee
We have used the fact that in those terms where $\phi c$ appears,  $c$ is not screened from another $c$ and so,
assuming $(c\phi)^2=0$, all these terms vanish.
Next we have moved $X$ from extreme left to extreme right using the cyclicity of the trace. We have done the same for $B$ and rearranged the terms so as to obtain the third expression above.

Now we will try to write $\langle \psi_L \psi_L \psi_L \rangle=\langle \psi_R \psi_R \psi_R \rangle $ in L-R symmetric form. To this end we write
\be
BcKc= \frac 12 (BcKc +cKcB) +\frac 12 [K,c]. \label{LR2}
\ee
Then
\be
&&\langle \psi_R \psi_R \psi_R \rangle=\0\\
&=&-\frac 18 \langle \sqrt{X}  (BcKc +cKcB)H\sqrt{X} \,\, \sqrt{X}  (BcKc +cKcB)H\sqrt{X}\,\,
 \sqrt{X}  (BcKc +cKcB)H\sqrt{X} \rangle\0\\
&=& -\frac 18 \langle[K,c]HX\,\, [K,c]HX\,\,[K,c]HX\rangle\0\\
&=& -\frac 38 \langle BcKcHX\,\, [K,c]HX\,\,[K,c]HX\rangle\0\\
&=& -\frac 38 \langle cKcBHX\,\, [K,c]HX\,\,[K,c]HX\rangle\0\\
&=& -\frac 38 \langle BcKc HX \,\, BcKc HX \,\, [K,c]HX\rangle\0\\
&=& -\frac 38 \langle cKcB HX \,\, cKcB HX \,\, [K,c] HX\rangle\label{LR3}
\ee
In the last two lines we have used the collision of two $B$'s (and the cyclicity property) to get rid of two additional terms.\\

\no The RHS expression of the first line is already in L-R symmetric form, so we leave it there. Now consider $\langle[K,c]HX\,\, [K,c]HX\,\,cKcB HX\rangle$. We can move $B$ around using cyclicity and (\ref{LR0}) and get
\be
&&\langle[K,c]HX\,\, [K,c]HX\,\,cKcB HX\rangle=\langle[K,c]HX\,\, [K,c]HX\,\,BcKc HX\rangle=\0\\
 &&\langle[K,c]HX\,\, [K,c]HX\,\,(cKcB+[K,c]) HX\rangle\label{LR4}
\ee
This means in particular that
\be
\langle[K,c]HX\,\, [K,c]HX\,\,[K,c] HX\rangle=0\label{LR5}
\ee
So the second line in the RHS of (\ref{LR3}) vanishes. Let us consider next the fourth and last line of (\ref{LR3})
\be
&&\langle cKcBHX\,\, [K,c]HX\,\,[K,c]HX\rangle+ \langle cKcB HX \,\, cKcB HX \,\, [K,c] HX\rangle\0\\
&& \langle cKcBHX\,\, [K,c]HX\,\,[K,c]HX\rangle+ \langle cKc HX \,\,B cKcB HX \,\, [K,c] HX\rangle\0\\
&&  \langle cKcBHX\,\, [K,c]HX\,\,[K,c]HX\rangle+ \langle cKc HX \,\,[K,c]B HX \,\, [K,c] HX\rangle\0\\
&&  \langle cKcBHX\,\, [K,c]HX\,\,[K,c]HX\rangle-\langle cKc BHX \,\,[K,c] HX \,\, [K,c] HX\rangle=0.\label{LR6}
\ee
What remains to be considered is
\be
&&\langle BcKcHX\,\, [K,c]HX\,\,[K,c]HX\rangle+\langle BcKc HX \,\, BcKc HX \,\, [K,c]HX\rangle\0\\
&&=\langle BcKcHX\,\, [K,c]HX\,\,[K,c]HX\rangle- \langle cKc HX \,\, BcKcB HX \,\, [K,c]HX\rangle\0\\
&&=\langle BcKcHX\,\, [K,c]HX\,\,[K,c]HX\rangle- \langle cKc HX \,\, [K,c]BHX \,\, [K,c]HX\rangle\0\\
&&=\langle BcKcHX\,\, [K,c]HX\,\,[K,c]HX\rangle+ \langle cKcB HX \,\, [K,c]HX \,\, [K,c]HX\rangle\0\\
&& =\langle \{B,cKc\} HX\,\, [K,c]HX\,\,[K,c]HX\rangle\0\\
&&=-  \langle[K,c]HX\,\, \{B,cKc\} HX\,\, [K,c]HX\,\,\rangle.\label{LR7}
\ee
The last expression can be written as
\be
- \langle[K,c]HX\,\, \{BcKc,cKcB\} HX\,\, [K,c]HX\,\,\rangle= - \langle\sqrt{X} [K,c]HX\,\, \{B,cKc\} HX\,\, [K,c]H\sqrt{X}\,\,\rangle\0\label{LR8}
\ee
and it is L-R symmetric.

Finally
\be
&&-\langle \psi_R \psi_R \psi_R \rangle=\0\\
&=&  \frac 18 \langle \sqrt{X}  (BcKc +cKcB)H\sqrt{X} \,\, \sqrt{X}  (BcKc +cKcB)H\sqrt{X}\,\,
 \sqrt{X}  (BcKc +cKcB)H\sqrt{X} \rangle\0\\
&& + \langle\sqrt{X} [K,c]H\sqrt{X}\,\,\sqrt{X} \{B,cKc\} H\sqrt{X}\,\, \sqrt{X}[K,c]H\sqrt{X}\,\,\rangle\label{LRenergy}
\ee
which is LR symmetric, and therefore real.

\section{Cubic Superstring generalization}

As it was the case for the Tachyon Vacuum,~\cite{Erler:2007xt}, our
lump solution can also be easily generalized to a corresponding
solution in the cubic SSFT. The solution has the same pure gauge
form as in the bosonic case, but now $Q$ acts differently. In
particular \be Qc&=&cKc-\gamma^2. \ee The BRST variation of the seed
$\phi(X)$ is given by \be [Q,\phi(w)]=c\partial\phi(w)+\partial
c\delta\phi(w)+\gamma\,(\psi\delta'\phi)(w), \ee

\be
Q\phi=c[K,\phi]+[K, c]\delta\phi(w)+\gamma(\psi\delta'\phi).
\ee
Assuming that the seed $\phi$ is solely made of the $X$ field we have
\be
\delta\phi(X)&=&-\del^2_X\phi(X)\\
\delta'\phi(X)&=&-\del_X\phi(X).
\ee
The singular gauge transformation which generates the solution is the same as the bosonic one
\be
V_L&=&1-\frac{1}{K+\phi}\phi Bc\\
V_L^{-1}&=&1+\frac{1}{K}\phi Bc.
\ee

Due to the different action of the BRST charge the solution will contain superstring corrections
\be
\psi_L=V_L Q V_L^{-1}&=&c\phi-\frac1{K+\phi}BQ(c\phi)\\
&=&c\phi-\frac1{K+\phi}\left[(\phi-\delta\phi)Bc\del c+(\psi\cdot\delta'\phi)Bc\gamma-B\gamma^2\phi\right].
\ee
Again, the cohomology of the solution can be formally trivialized (the superstring corrections add up to zero)
\be
{\cal Q}_{\psi_L}\frac B{K+\phi}=1,
\ee
so a singular $\frac1{K+\phi}$ is needed to have a solution in a
 different gauge orbit from the superstring `tachyon vacuum' of \cite{Erler:2007xt}.\\
In this case, however, the regularity of the solution involves
different `sectors' labeled by their ghost structure. It would be
interesting to understand if one can find explicit examples of seeds
giving rise to solutions different from the tachyon
vacuum. Notice that the solution is
in the GSO(+) sector so, if it can be made regular, it would
probably describe a codimension 1 `brane' in the GSO(+) sector,
which does not seem to exist in string theory. On the other hand,
the impossibility of building a regular solution (with a seed in the
GSO(+) sector) would be consistent with the spectrum of string
theory. We think this is an interesting problem which can contribute
to the long debate about the consistency of the cubic RNS
theory~\cite{Kroyter:2009zj, Kroyter:2009bg, Kroyter:2009rn,
Aref'eva:2008ad,  Aref'eva:2010yd,  Erler:2010pr}.

\section{Computations for the  $:X^2:$ deformation}

In order to use Witten's results, \cite{Witten92},  we have to map
the cylinder  to the unit disk. Given a cylinder of width $T$ we
first scale it to a canonical cylinder of width 1 and obtain
\be
\left\langle e^{-\int_0^T ds\, [u(X^2(s)+2\log
u+2A)]}\right\rangle_{C_T}=\left\langle e^{-\int_0^1 ds\,
[Tu(X^2(s)+2\log Tu+2A)]}\right\rangle_{C_1}.
\ee
Now we can map to
the unit disk with
\be
 w=-e^{2\pi i \tilde z},
 \ee
 where $w$ is the
global coordinate on the unit disk and $\tilde z$ the global
coordinate on the canonical cylinder of width 1. Explicitly we get
\be
g_A(u)\equiv \left\langle e^{-\int_0^1 ds\, [u(X^2(s)+2\log
u+2A)]}\right\rangle_{C_1} =\left\langle e^{-\int_0^{2\pi} d\theta\,
\left[\frac u{2\pi}\left(X^2(e^{i\theta}) +2\log \frac
u{2\pi}+2A\right)\right]}\right\rangle_{Disk}.\label{gu}
\ee
Now to
explicitly evaluate \eqref{gu} we use the result of \cite{Witten92},
keeping in mind that in there $\alpha'=2$ while here we use the more
common $\alpha'=1$. Thus we have
 \be
 Z(u)\equiv\left\langle e^{-\int_0^{2\pi}
d\theta\, \frac
u{4\pi}X^2(e^{i\theta})}\right\rangle_{Disk}=K\,\sqrt{u}\exp(\gamma
u)\Gamma(u)\label{Zu}.
\ee
Here $K$ is a $u$--independent
normalization constant, which depends on the way one normalizes the
zero mode $x$ integration. We normalize it as follows:
$$\int_0^\infty\frac{dx}{2\pi}=\bra 0\ket 0=\frac{V_1}{2\pi}.$$
So, in our conventions, we have
\be
 K=\frac1{\sqrt{2\pi}}.
\ee
Therefore, the partition function on canonical cylinder in
\eqref{gu} is
 \be
 \left\langle e^{-\int_0^1\,\phi_u(s)
ds}\right\rangle_{C_1}
=\frac1{\sqrt{2\pi}}\sqrt{2u}\,\Gamma(2u)e^{u\left(2\gamma-2\log\frac
{u}{2\pi}-2A\right)}.
\ee

Now we have to determine the $u$--independent number $A$ in order
for the  partition function to be finite in the $u\to\infty$ limit.
Using Stirling approximation we have
\be
\sqrt{u}\,\Gamma(u)\,e^{-u\log
u}\approx_{u\to\infty}\sqrt{2\pi}\,e^{-u},
\ee
which implies
\be
\lim_{u\to\infty}\left\langle e^{-\int_0^1\,\phi_u(s)
ds}\right\rangle_{C_1}=e^{2u(\gamma-1+\log4\pi-A)}.
\ee
The only way
this can be finite in the large $u$ limit is
\be
A=\gamma-1+\log4\pi.
\ee
 Any other choice of $A$ would give a
divergent or vanishing partition function in the infrared. This
result is a little bit different from \cite{Ellwood:2009zf} by the
 $\log4\pi$ term, because in \cite{Ellwood:2009zf} of some different conventions.

With this unique choice of $A$ we get
\be
g(u)\equiv\left\langle e^{-\int_0^1\,\phi_u(s) ds}\right\rangle_{C_1}
&=&\frac1{\sqrt{2\pi}} \sqrt{2u}\,\Gamma(2u)\left(\frac e{2u}\right)^{2u}\\
\left\langle e^{-\int_0^T\,\phi_u(s) ds}\right\rangle_{C_T}&=&g(Tu).
\ee
In the deep IR we get a properly normalized  partition function for Dirichlet boundary
conditions
 \be
\lim_{u\to\infty}g(u)=K\sqrt{2\pi}=\frac1{\sqrt{2\pi}}\sqrt{2\pi}=1,
\ee
while in the UV (no deformation present) we get the partition
function with Neumann boundary condition, which is the BCFT on which
the theory is defined from the start
\be
\lim_{u\to0}g(u)=\lim_{u\to0}\frac1{2\sqrt{\pi u}}=
 \delta(0)=\int_0^\infty\frac{dx}{2\pi}=\frac
{V_1}{2\pi}=\bra0\ket0_{SL(2,R)}. \ee This last divergence is
expected because the $X^2$ deformation cannot be defined on a
compact direction and all computations are done at infinite
transverse volume. The vacuum to vacuum amplitude is indeed
divergent ($\delta(p=0)$) because of the non compact zero mode of
$X$.

\subsection{Explicit results for one-point function}

In addition to \eqref{Zu}, from \cite{Witten92}, we can also find
the following one-point function \be \left\langle X^2
e^{-\int_0^{2\pi}d\theta\,\frac u{4\pi}X^2(\theta)}
\right\rangle_{Disk}&=& -2\del_u
Z(u)=\left(\frac1u-2H(u)\right)Z(u).
\ee
where $H(z)\equiv H_z$ is
the $z^{th}$ Harmonic Number. They can be defined as \be H(z)\equiv
H_z\equiv \gamma+\psi(z+1). \ee It inherits from the digamma
$(\psi(z))$ function the following property \be H(z)=\frac1z+H(z-1).
\ee We can then properly map this one point function to the cylinder
$C_T$ so as to evaluate the one point function of the
insertion $\phi_u(z)= u\left(X^2(z)+2\log u+2A\right)$ on the
cylinder. It is given by
 \be
\left\langle\phi_u e^{-\int_0^{T}ds\,\phi_u(s)}
\right\rangle_{C_T}&=&2ug_A(Tu)\left(\frac1{4Tu}-H(2Tu)
+\log\frac{Tu}{2\pi}+A\right),\ee where $A$ is  determined as in the previous subsection.
A non trivial consistency check of the chain of mappings and
conventions is given by
\be
\left\langle u\del_u\phi_u e^{-\int_0^{T}ds\,\phi_u(s)} \right\rangle_{C_T}&=&\left\langle(\phi_u+2u) e^{-\int_0^{T}ds\,\phi_u(s)} \right\rangle_{C_T}\label{onep}\\
&=&2ug_A(Tu)\left(\frac1{4Tu}-H(2Tu)+\log\frac{Tu}{2\pi}+A+1\right)
=-\frac u T \del_u g_A(Tu),\0
\ee where we have used the
identity
 \be \frac{-\frac u T \del_u g_A(Tu)}{2ug_A(Tu)}
=\left(\frac1{4Tu}-H(2Tu)+\log\frac{Tu}{2\pi}+A+1\right),\label{IDD}
\ee which can easily be verified by the use of the
properties of Harmonic Numbers or Digamma functions. We realize
that \eqref{onep} is the expression of the boundary changing
mechanism at work in section 4
\be \left\langle(\phi_u+2u)
e^{-\int_0^{T}ds\,\phi_u(s)} \right\rangle_{C_T}&=
&\left\langle(u\del_u\phi_u(s')) e^{-\int_0^{T}ds\,\phi_u(s)}
\right\rangle_{C_T}\0\\
&=&-\frac u T \del_u\left\langle e^{-\int_0^{T}ds\,\phi_u(s)}
\right\rangle_{C_T}. \ee

\subsection{Three--point functions: energy}

In section 6 we have obtained a general expression of the cubic term in the action which
involves a three-point function of $\del_u\phi_u$. In this
subsection we evaluate it for the solution generated by the
insertion $\phi_u(z)= u\left(X^2(z)+2\log u+2A\right)$. We will make
use of the Green's function in the presence of the boundary
deformation $\int_0^{2\pi}d\theta\,\frac u{4\pi\alpha'}X^2(\theta)$,
which was derived in \cite{Witten92}. On the disc it is given by
\be
G_{u}(z-w)&=&\langle X(z,\bar z)X(w,\bar
w)\rangle^{(u)}_{Disk}\0\\
&=&-\frac{\alpha'}2 \Big(\ln|z-w|^2+\ln|1-z\bar
w|^2-\frac 2u+2u\sum_{k=1}^{\infty}\frac{(z\bar w)^k+(\bar z
w)^k}{k(k+u)}\Big),\0
\ee
where we denoted the deformed correlator
\be
\langle(...)\rangle^{(u)}_{Disk}\equiv\langle(...)e^{\int_0^{2\pi}d\theta\,\frac u{4\pi\alpha'}X^2(\theta)}\rangle_{Disk}.
\ee
On the boundary of the disc
($z=e^{i\theta}$) we have
\be
G_{u}(\theta-\theta')&=&\sum_{k=-\infty}^{\infty}\frac{e^{ik(\theta-\theta')}}
{|k|+u}=\frac1u +2\sum_{k=1}^{\infty}\frac1{k+u}\cos
k(\theta-\theta'),\0\\
&=&\frac1u+\frac2{1+u} Re\left[e^{i(\theta-\theta')}\,_2\!F_1\left(1+u,1,2+u;e^{i(\theta-\theta')}\right)\right]
\ee
where we have used $\alpha'=1$, and $_2\!F_1$ is the hypergeometric function
\be
_2\!F_1(a,b,c;z)&=&_2\!F_1(b,a,c;z)=\sum_{k=0}^{\infty}\frac{(a)_k(b)_k}{(c)_k}\frac{z^k}{k!}\0\\
(x)_k&\equiv &\frac{\Gamma(x+k)}{\Gamma(x)}.
\ee

 Using this
Green's function and the Wick's theorem, in addition to the 1-point
function given in the previous section, we obtain the following
two-- and three--point functions for $X^2(\theta)$, \be
&&\left\langle X^2(\theta)
\right\rangle^{(u)}_{Disk}=Z(u)\Big(\frac1u-2H(u)\Big)\equiv
Z(u)h_u,\0 \\
&& \left\langle X^2(\theta) X^2(\theta')
\right\rangle^{(u)}_{Disk}=Z(u)\Big(2G_{u}^2(\theta-\theta')+h_u^2\Big)\0\\
&& \left\langle X^2(\theta_1) X^2(\theta_2)X^2(\theta_3)
\right\rangle^{(u)}_{Disk}=Z(u)\Big\{8G_{u}(\theta_1-\theta_2)G_{u}
(\theta_1-\theta_3)G_{u}(\theta_2-\theta_3)\0\\
&&~~~~~~~~~~~~~~~~~~~~~~+2h_u\Big(G_{u}^2(\theta_1-\theta_2)
+G_{u}^2(\theta_1-\theta_3)+G_{u}^2(\theta_2-\theta_3)
\Big)+h_u^3\Big\}\label{threeP} \ee

The cubic term is proportional to
\be &&\langle \psi_u
\psi_u\psi_u\rangle = - \Big\langle \frac 1{K+{\phi}_u}
({\phi}_u+2u) BcKc
 \frac 1{K+{\phi}_u} ({\phi}_u+2u) BcKc  \frac1{K+{\phi}_u} ({\phi}_u+2u) BcKc
 \Big\rangle\0\0\\
 &&=-\int_{0}^\infty dt_1dt_2dt_3{\cal E}_0(t_1,t_2,t_3)\Big\langle
 ({\phi}_u(t_1+t_2)+2u)
 ({\phi}_u(t_1)+2u)({\phi}_u(0)+2u)e^{-\int_0^Tds{\phi}_u(s)}
 \Big\rangle_{C_T}\0
 \ee
 where ${\cal E}_0(t_1,t_2,t_3)$ is the contribution of the ghost
 sector and $T=t_1+t_2+t_3$.
 To use \eqref{threeP} we have to map the $C_T$ to the unit disc by
 \be
 w(z)=-e^{\frac{2\pi iz}{T}}.
 \ee
 Then we get,
 \be \langle \psi_u \psi_u\psi_u\rangle &=&-\int_{0}^\infty dt_1dt_2dt_3{\cal
E}_0(t_1,t_2,t_3)e^{-2uT\big(\ln(\frac
{uT}{2\pi})+A\big)}\0\\
&\times& u^3\Big\langle
\Big(X^2(\theta_{t_1+t_2})+2\big(\ln(\frac{uT}{2\pi})+A+1\big)\Big)
 \Big(X^2(\theta_{t_1})+2\big(\ln(\frac{uT}{2\pi})+A+1\big)\Big)\0\\
 &\times&\Big(X^2(0)+2\big(\ln(\frac{uT}{2\pi})+A+1\big)
 \Big)e^{-\int_0^{2\pi}d\theta
 \frac{2uT}{4\pi}X^2(\theta)}\Big\rangle_{Disk},
 \ee
 where $\theta_{t}=\frac{2\pi t}T$.
 Using \eqref{threeP}, setting $A=\gamma-1+\ln4\pi$ and simplifying,
 \be
 \langle \psi_u
\psi_u\psi_u\rangle &=&-\int_{0}^\infty dt_1dt_2dt_3{\cal
E}_0(t_1,t_2,t_3)u^3e^{-2uT\big(\ln(
{2uT})+\gamma-1\big)}Z(2uT)\0\\
&\cdot&\Bigg\{8\Big(\frac{h_{2uT}}2+\ln(2uT)+\gamma\Big)^3\0\\
&+&4\Big(\frac{h_{2uT}}2+\ln(2uT)+\gamma\Big)\Big(G_{2uT}^2(\frac{2\pi
t_1}T)+G_{2uT}^2(\frac{2\pi (t_1+t_2)}T)+G_{2uT}^2(\frac{2\pi
t_2}T)\Big)\0\\
&+&8G_{2uT}(\frac{2\pi t_1}T)G_{2uT}(\frac{2\pi
(t_1+t_2)}T)G_{2uT}(\frac{2\pi t_2}T)\Bigg\}. \ee
 Applying the identity \eqref{IDD} we get
\be \langle \psi_u \psi_u\psi_u\rangle &=&-\int_{0}^\infty
dt_1dt_2dt_3{\cal
E}_0(t_1,t_2,t_3)u^3g(uT)\Bigg\{8\Big(-\frac1{2}\frac{\partial_{uT}g(uT)}{g(uT)}\Big)^3\0\\
&+&4\Big(-\frac1{2}\frac{\partial_{uT}g(uT)}{g(uT)}\Big)\Big(G_{2uT}^2(\frac{2\pi
t_1}T)+G_{2uT}^2(\frac{2\pi (t_1+t_2)}T)+G_{2uT}^2(\frac{2\pi
t_2}T)\Big)\0\\
&+&8G_{2uT}(\frac{2\pi t_1}T)G_{2uT}(\frac{2\pi
(t_1+t_2)}T)G_{2uT}(\frac{2\pi t_2}T)\Bigg\}\label{LLL} \ee

In section 6 we have shown that the above quantity is $u$ independent. We can
 easily confirm this  by doing a convenient
change of variables $(t_1,t_2,t_3)\to (T,x,y)$, where
\be
x&=&\frac{t_1}{T}\\
y&=&\frac{t_2}{T}.
\ee
The matter part, (\ref{LLL}), (before
integration) can be written as
$$u^3 F(uT,x,y),$$
 where
 \be
  F(uT,x,y)=&g(uT)\Bigg\{8\Big(-\frac1{2}\frac{\partial_{uT}g(uT)}{g(uT)}\Big)^3
  +8G_{2uT}(2\pi x)G_{2uT}(2\pi(x+y))G_{2uT}(2\pi y)\0\\
+&4\Big(-\frac1{2}\frac{\partial_{uT}g(uT)}{g(uT)}\Big)\Big(G_{2uT}^2(2\pi
x)+G_{2uT}^2(2\pi(x+y))+G_{2uT}^2(2\pi y)\Big) \Bigg\}.
\ee
 Computing the ghost correlator we have
\be
\E_0(t_1,t_2,t_3)=\left\langle Bc\del c(t_1+t_2)\del c(t_1) \del
c(0) \right\rangle_{C_T}=\E(x,y)=-\frac4\pi \,\sin\pi x\,\sin\pi y\,
\sin\pi(x+y)\0,
\ee
notice that the ghost correlator only depends on
$x$ and $y$, which are scale invariant coordinates.\\
We now change variables of integrations,
\be
\int_0^\infty
dt_1 \int_0^\infty dt_2\int_0^\infty dt_3=\int_0^\infty dT\;
T^2\int_0^1 dx\int_0^{1-x} dy,
\ee
so, collecting everything, we obtain
\be
-\frac16Tr[\psi_u\psi_u\psi_u]
&=&\frac16\int_0^\infty dT\; T^2\int_0^1
dx\int_0^{1-x} dy\,\E(x,y)\,u^3 F(uT,x,y).
\ee
As a last step we do
the obvious change of coordinate
\be t&=&uT, \ee
 which readily gives
\be -\frac16Tr[\psi_u\psi_u\psi_u]=\frac16\int_0^\infty dt\; t^2\int_0^1
dx\int_0^{1-x} dy\,\E(x,y)\, F(t,x,y).\label{last}
\ee
 Notice that $u$ has
completely disappeared, so the energy is $u$--independent. In the UV region $t\to0$, the integrand goes like $1/t^{3/2}\sim \del_t1/\sqrt{t}$, which is the expected volume divergence. As far as the IR ($t\to\infty$) is concerned, the finiteness of (\ref{last}) has been proven in \cite{BGT}.

\section{Ordered exponential insertion}

In this appendix we show how to write down the insertion $\frac 1{K+\phi}$ inside a correlator
by a Schwinger--like representation. Let us start from the parametric representation
\be
\frac 1{K+\phi}= \int_0^\infty dt\,e^{-t(K+\phi)}\label{par}
\ee
To make sense of this formula we need to split the exponential. The RHS can be understood as the action of the operator
$K'\sim K^L_1$ and the field operator $\phi'$ on the identity string field,
\be
\int_0^\infty dt\,e^{-t(K+\phi)}= \int_0^\infty dt\,e^{-t(K'+\phi')}|I\rangle\label{par1}
\ee
We will split the operator $e^{-t(K'+\phi')}$ in two exponentials. To this end we use the Zassenhaus formula, which holds for finite matrices,
\be
e^{t(X+Y)} &=& e^{tX} e^{tY} e^{-\frac {t^2}2 [X,Y] }
e^{\frac {t^3}{3!}(2[Y,[X,Y]]+[X,[X,Y]])}\,e^{- \frac {t^4}{4!}([[[X,Y],X],X]+3[[[X,Y],X],Y]+3[[[X,Y],Y],Y])}\ldots,\0
\ee
and extend it to general operators.

Choosing $X=-K', Y=-\phi'$, inside a correlator we can assume $[[X,Y],Y]=0$, $[[[X,Y],X],Y]=0$, etc. Therefore the formula simplifies to
\be
e^{-t(K'+\phi')}= e^{-tK'} e^{-S(t)},\label{zassen1}
\ee
where
\be
S(t)=t\phi'+ \frac {t^2}2 [K',\phi']+ \frac {t^3}{3!} [K',[K',\phi']] +
 \frac {t^4}{4!}[K',[K',[K',\phi']]]+\ldots, \label{St}
\ee
with $S(0)=0$. Now, $\phi'$ denotes the field $\phi'$ evaluated at some point $t_0$. Thus
\be
\d_t S(t) &=& \phi'(t_0) + t [K',\phi'(t_0)]+ \frac {t^2}{2!} [K',[K',\phi'(t_0)]] +
 \frac {t^3}{3!}[K',[K',[K',\phi'(t_0)]]]+\ldots\0\\
 &=& \phi'(t+t_0)\label{St1}
\ee
and
\be
S(t)=  \int_{t_0}^{t+t_0} ds\, \phi'(s)\label{St2}
\ee
Applying back (\ref{zassen1}) to the identity string field, we obtain

\be
\frac 1{K+\phi}= \int_0^\infty dt\, e^{-tK} e^{-\int_{t_0}^{t+t_0} ds \phi(s)}
\label{parfin}
\ee

\end{document}